\documentclass{aastex63}
\usepackage{xcolor}
\usepackage{multirow}
\usepackage[utf8]{inputenc}
\usepackage{textcomp}

\definecolor{pink}{rgb}{1,0.1,.6}

\newcommand{\lco}{LCOGT}

\shorttitle{Two Warm Super-Earths Transiting the Nearby M Dwarf TOI-2095}
\shortauthors{Quintana et al.}

\begin{document}

\title{Two Warm Super-Earths Transiting the Nearby M Dwarf, TOI-2095}

\correspondingauthor{Elisa V. Quintana}
\email{elisa.quintana@nasa.gov}

\author[0000-0003-1309-2904]{Elisa V. Quintana}
\affiliation{NASA Goddard Space Flight Center, Greenbelt, MD 20771, USA}

\author[0000-0002-0388-8004]{Emily A. Gilbert}
\affiliation{Jet Propulsion Laboratory, California Institute of Technology, Pasadena, CA 91109, USA}

\author[0000-0001-7139-2724]{Thomas Barclay}
\affiliation{NASA Goddard Space Flight Center, Greenbelt, MD 20771, USA}
\affiliation{University of Maryland, Baltimore County, 1000 Hilltop Cir, Baltimore, MD 21250, USA}

\author[0000-0003-2565-7909]{Michele L. Silverstein}
\affiliation{University of Maryland, Baltimore County, 1000 Hilltop Cir, Baltimore, MD 21250, USA}
\affiliation{NASA Goddard Space Flight Center, Greenbelt, MD 20771, USA}

\author[0000-0001-5347-7062]{Joshua E. Schlieder}
\affiliation{NASA Goddard Space Flight Center, Greenbelt, MD 20771, USA}

\author[0000-0001-5383-9393]{Ryan Cloutier}
\affiliation{Department of Physics \& Astronomy, McMaster University, 1280 Main St West, Hamilton, ON, L8S 4L8, Canada}

\author[0000-0002-8964-8377]{Samuel N. Quinn}
\affiliation{Center for Astrophysics $\mid$ Harvard \& Smithsonian, 60 Garden St, Cambridge, MA, 02138, USA}

\author[0000-0001-8812-0565]{Joseph E. Rodriguez}
\affiliation{Center for Data Intensive and Time Domain Astronomy, Department of Physics and Astronomy, Michigan State University, East Lansing, MI 48824, USA}

\author[0000-0001-7246-5438]{Andrew Vanderburg}
 \affiliation{Department of Physics and Kavli Institute for Astrophysics and Space Research, Massachusetts Institute of Technology, Cambridge, MA, 02139, USA}
 
\author[0000-0001-5084-4269]{Benjamin J. Hord}
\affiliation{Astronomy Department, University of Maryland, College Park, MD 20742, USA}
\affiliation{NASA Goddard Space Flight Center, Greenbelt, MD 20771, USA}

\author[0000-0002-2457-272X]{Dana R. Louie}
\altaffiliation{NASA Postdoctoral Program Fellow}
\affiliation{NASA Goddard Space Flight Center, Greenbelt, MD 20771, USA}

\author[0000-0002-7084-0529]{Colby Ostberg}
\affiliation{Department of Earth and Planetary Sciences, University of California, Riverside, CA 92521, USA}

\author[0000-0002-7084-0529]{Stephen R. Kane}
\affiliation{Department of Earth and Planetary Sciences, University of California, Riverside, CA 92521, USA}

\author[0000-0001-6541-0754]{Kelsey Hoffman} 
\affiliation{SETI Institute, 189 Bernardo Ave, Suite 200, Mountain View, CA 94043, USA}
\affiliation{Bishop's University, Department of Physics and Astronomy, 2600 College Street, Sherbrooke, QC, J1M 1Z7, Canada}

\author[0000-0002-5904-1865]{Jason F. Rowe}
\affiliation{Bishops University, 2600 College St, Sherbrooke, QC J1M 1Z7, Canada}

\author[0000-0001-6285-267X]{Giada N. Arney}
\affiliation{NASA Goddard Space Flight Center, Greenbelt, MD 20771, USA}

\author{Prabal Saxena}
\affiliation{NASA Goddard Space Flight Center, Greenbelt, MD 20771, USA}

\author{Taran Richardson}
\affiliation{Department of Physics and Astronomy, Howard University, 2400 6th St NW, Washington, DC 20059, USA}

\author[0000-0001-8933-6878]{Matthew S. Clement}
\affiliation{Johns Hopkins APL, 11100 Johns Hopkins Rd, Laurel, MD 20723, USA}
\affiliation{Earth and Planets Laboratory, Carnegie Institution for Science, 5241 Broad Branch Road, NW, Washington, DC 20015, USA}

\author{Nicholas M. Kartvedt} %intern
\affiliation{Department of Physics, United States Naval Academy, 572C Holloway Rd., Annapolis, MD 21402, USA}

\author[0000-0002-8167-1767]{Fred C. Adams}
\affiliation{Department of Physics, University of Michigan, 450 Church St, Ann Arbor, MI 48109, USA}

\author{Marcus Alfred}
\affiliation{Department of Physics and Astronomy, Howard University, 2400 6th St NW, Washington, DC 20059, USA}

\author[0000-0002-2580-3614]{Travis Berger}
\affiliation{NASA Goddard Space Flight Center, Greenbelt, MD 20771, USA}

\author[0000-0001-6637-5401]{Allyson Bieryla}
\affiliation{Center for Astrophysics $\mid$ Harvard \& Smithsonian, 60 Garden St, Cambridge, MA, 02138, USA}

\author{Paul Bonney}
\affiliation{Department of Physics, University of Arkansas, 1 University of Arkansas, Fayetteville, AR 72701, USA}

\author{Patricia Boyd}
\affiliation{NASA Goddard Space Flight Center, Greenbelt, MD 20771, USA}

\author[0000-0001-9291-5555]{Charles Cadieux}
\affiliation{Universit\'e de Montr\'eal, D\'epartement de Physique, IREX, Montr\'eal, QC H3C 3J7, Canada}

\author[0000-0003-1963-9616]{Douglas Caldwell}
\affiliation{NASA Ames Research Center, Moffett Field, CA 94035, USA}

\author[0000-0002-5741-3047]{David R. Ciardi}
\affiliation{Caltech IPAC – NASA Exoplanet Science Institute 1200 E. California Ave, Pasadena, CA 91125, USA}

\author[0000-0002-9003-484X]{David Charbonneau}
\affiliation{Center for Astrophysics $\mid$ Harvard \& Smithsonian, 60 Garden Street, Cambridge, MA 02138, USA}

\author[0000-0001-6588-9574]{Karen A.\ Collins}
\affiliation{Center for Astrophysics $\mid$ Harvard \& Smithsonian, 60 Garden St, Cambridge, MA, 02138, USA}

\author[0000-0001-8020-7121]{Knicole D. Col\'on}
\affiliation{NASA Goddard Space Flight Center, Greenbelt, MD 20771, USA}

\author[0000-0003-2239-0567]{Dennis M.\ Conti}
\affiliation{American Association of Variable Star Observers, 185 Alewife Brook Parkway, Suite 410, Cambridge, MA 02138, USA}

\author{Mario Di Sora}
\affiliation{Campo Catino Astronomical Observatory, Regione Lazio, Guarcino (FR), 03010 Italy}

\author[0000-0001-6285-267X]{Shawn Domagal-Goldman}
\affiliation{NASA Goddard Space Flight Center, Greenbelt, MD 20771, USA}

\author[0000-0003-4206-5649]{Jessie Dotson}
\affiliation{NASA Ames Research Center, Moffett Field, CA 94035, USA}

\author{Thomas Fauchez}
\affiliation{NASA Goddard Space Flight Center, Greenbelt, MD 20771, USA}

\author[0000-0002-3164-9086]{Maximilian N. Günther}
\affiliation{European Space Agency (ESA), European Space Research and Technology Centre (ESTEC), Keplerlaan 1, 2201 AZ Noordwijk, The Netherlands}

\author[0000-0002-3385-8391]{Christina Hedges}
\affiliation{NASA Goddard Space Flight Center, Greenbelt, MD 20771, USA}
\affiliation{University of Maryland, Baltimore County, 1000 Hilltop Cir, Baltimore, MD 21250, USA}

\author{Giovanni Isopi}
\affiliation{Campo Catino Astronomical Observatory, Regione Lazio, Guarcino (FR), 03010 Italy}

\author{Erika Kohler}
\affiliation{NASA Goddard Space Flight Center, Greenbelt, MD 20771, USA}

\author{Ravi Kopparapu}
\affiliation{NASA Goddard Space Flight Center, Greenbelt, MD 20771, USA}

\author[0000-0001-5347-7062]{Veselin B. Kostov}
\affiliation{NASA Goddard Space Flight Center, Greenbelt, MD 20771, USA}

\author{Jeffrey A. Larsen} %larsen@usna.edu
\affiliation{Department of Physics, United States Naval Academy, 572C Holloway Rd., Annapolis, MD 21402, USA}

\author{Eric Lopez}
\affiliation{NASA Goddard Space Flight Center, Greenbelt, MD 20771, USA}

\author{Franco Mallia}
\affiliation{Campo Catino Astronomical Observatory, Regione Lazio, Guarcino (FR), 03010 Italy}

\author{Avi Mandell}
\affiliation{NASA Goddard Space Flight Center, Greenbelt, MD 20771, USA}

\author[0000-0001-7106-4683]{Susan E. Mullally}
\affiliation{Space Telescope Science Institute, 3700 San Martin Drive, Baltimore, MD, 21218, USA}

\author[0000-0002-8090-3570]{Rishi R. Paudel}
\affiliation{NASA Goddard Space Flight Center, Greenbelt, MD 20771, USA}
\affiliation{University of Maryland, Baltimore County, 1000 Hilltop Cir, Baltimore, MD 21250, USA}

\author[0000-0003-0501-2636]{Brian P. Powell}
\affiliation{NASA Goddard Space Flight Center, Greenbelt, MD 20771, USA}

\author{George R. Ricker}
\affiliation{Department of Physics and Kavli Institute for Astrophysics and Space Research, Massachusetts Institute of Technology, Cambridge, MA, 02139, USA}

\author{Boris S. Safonov}
\affiliation{Sternberg Astronomical Institute, Lomonosov Moscow State University, 119992, Universitetskij prospekt 13, Moscow, Russia}

\author[0000-0001-8227-1020]{Richard P. Schwarz}
\affiliation{Center for Astrophysics \textbar \ Harvard \& Smithsonian, 60 Garden Street, Cambridge, MA 02138, USA}

\author[0000-0003-3904-6754]{Ramotholo Sefako} 
\affiliation{South African Astronomical Observatory, P.O. Box 9, Observatory, Cape Town 7935, South Africa}

\author[0000-0002-3481-9052]{Keivan G.\ Stassun}
\affiliation{Vanderbilt University, Department of Physics \& Astronomy, 6301 Stevenson Center Ln., Nashville, TN 37235, USA}
\affiliation{Fisk University, Department of Physics, 1000 18th Ave. N., Nashville, TN 37208, USA}

\author{Robert Wilson}
\affiliation{NASA Goddard Space Flight Center, Greenbelt, MD 20771, USA}

\author[0000-0002-4265-047X]{Joshua N.\ Winn}
\affiliation{Department of Astrophysical Sciences, Princeton University, Princeton, NJ 08544, USA}

\author[0000-0001-6763-6562]{Roland K. Vanderspek}
\affiliation{Department of Physics and Kavli Institute for Astrophysics and Space Research, Massachusetts Institute of Technology, Cambridge, MA 02139, USA}

\begin{abstract}
We report the detection and validation of two planets orbiting TOI-2095 (TIC 235678745). The host star is a 3700K M1V dwarf with a high proper motion. The star lies at a distance of 42 pc in a sparsely populated portion of the sky and is bright in the infrared (K = 9). With data from 24 Sectors of observation during TESS’s Cycles 2 and 4, TOI-2095 exhibits two sets of transits associated with super-Earth-sized planets. The planets have orbital periods of 17.7 days and 28.2 days and radii of 1.30 R$_\oplus$ and 1.39 R$_\oplus$, respectively. Archival data, preliminary follow-up observations, and vetting analyses support the planetary interpretation of the detected transit signals. The pair of planets have estimated equilibrium temperatures of approximately 400 K, with stellar insolations of 3.23 and 1.73 $S_\oplus$, placing them in the Venus zone. The planets also lie in a radius regime signaling the transition between rock-dominated and volatile-rich compositions. They are thus prime targets for follow-up mass measurements to better understand the properties of warm, transition radius planets. The relatively long orbital periods of these two planets provide crucial data that can help shed light on the processes that shape the composition of small planets orbiting M dwarfs.

\end{abstract}
% \section{Notes}

% % Target Name LSPM J1902+7525 in constellation Draco

\section{Introduction} \label{sec:intro} %\comment{Elisa/Em}

Amongst the most remarkable surprises revealed from large photometric transit surveys like Kepler/K2 \citep{Borucki2010,howell14} and the Transiting Exoplanet Survey Satellite (TESS) \citep{Ricker2015} is that (1) the most abundant type of planets are those with sizes in-between that of Earth and Neptune (1--4$\times$Earth's radius, R$_{\oplus}$), and (2) the size distribution of this population appears to exhibit a bimodal feature, with a dearth of planets near 1.5--2$\times$R$_{\oplus}$ for planets with orbital periods less than about 100 days \citep{Owen2013, Fulton2017, VanEylen2018}. The known planets smaller than Neptune, therefore, fall within two distinct populations: (1) enveloped terrestrial planets, or ``super-Earths'', that are likely rocky, and (2) small planets with extended H/He envelopes, or ``mini-Neptunes''. While the population analysis is primarily based on planets orbiting hotter stars, this bimodality has also been identified for exoplanets orbiting M dwarfs. However, in the case of M dwarfs, the center of the radius valley appears to shift to smaller planet sizes with decreasing stellar mass \citep{Cloutier2020}, albeit not without controversy \citep{Luque2022}. There has since been a large body of research aimed at understanding what formation mechanisms drive these two populations, and whether there are separate formation paths for planets orbiting low-mass stars. 

Several formation scenarios have been proposed to explain this bimodal distribution of the small exoplanet population. Because the aforementioned transit surveys are sensitive to planets orbiting close to their host stars (compared to the Earth-Sun system), the effect of the heating from the star may play a large role in the formation and retainment of exoplanet atmospheres, especially in relation to photoevaporation of atmospheres by high-energy (X-ray and extreme ultraviolet, XUV) radiation from the host star \citep{Owen2020}. This degradation of the atmosphere due to incident stellar illumination and atmosphere loss driven from energy generated in the planet's core \citep[i.e. core-powered mass loss;][]{Ginzburg2018} are mechanisms that may drive atmospheric escape on small exoplanets. The former is expected to occur early in the formation process, whereas the latter is expected to take place over billions of years \citep{Cloutier2020}. Other scenarios that can explain the presence or lack of an atmosphere include accretion or erosion of atmospheric material during the tenuous impact phases of planet formation \citep{Inamdar2015}, or the primordial presence and makeup of the gas component of protoplanetary disks while planets are forming \citep[e.g. rocky planets may simply be born in a gas-poor or gas-depleted environment;][]{Lee2021,Lopez2018}

Photoevaporation and core-powered mass loss are mechanisms that can explain atmosphere loss for planets orbiting low-mass stars only if their orbital periods are less than about 20--30 days \citep{Cloutier2020}. Planets found on wider orbits are thus valuable to test the different proposed mechanisms that sculpt the radius valley. Specifically, with follow-up mass (and therefore composition) measurements, if planets on wider orbits are shown to be rocky, then we can deduce that they were likely formed that way.

Recently, \citet{Luque2022} have suggested that there are not two, but three populations of planets orbiting close to M dwarfs: rocky, water-rich, and gas-rich. They argue that planet bulk compositions are not driven by mass loss but rather by where the planets originally formed, such as rocky planets forming close to the ice line, and lower-density water worlds forming beyond the ice line and migrating inward. They see no evidence of any stellar insolation or orbital period dependence on the planet's density which would be expected if the planetary composition was driven by mass loss. However, further research is needed to fully understand the origin and evolution of these types of planets \citep{Rogers2023}. Finding exoplanets that are amenable to mass measurements (which is the primary goal of TESS) will provide a way to test these theories, and help shed light on what truly shapes the compositions and atmospheres of small worlds.
 
Herein, we validate and characterize two planets orbiting the nearby M dwarf, TOI-2095, detected using TESS. The TOI-2095 system resides within the TESS and JWST continuous viewing zone (CVZ), and the host star is relatively bright (M1V) and quiet, making both planets excellent targets for obtaining follow-up precision radial velocity (mass) measurements. These planets have relatively long orbital periods (17.7 and 28.2 days), placing them in a regime where their atmospheres are less vulnerable to photoevaporation. Future mass measurements can therefore provide constraints on the formation and evolution processes that shape the composition of small planets orbiting M dwarfs.

\section{TESS Observations}
\label{sec:TESS} 

TOI-2095 is an M dwarf that resides in the Northern TESS Continuous Viewing Zone (CVZ), and Table \ref{tab:sectors} shows the specific observations taken by TESS. TOI-2095 was observed in all 13 sectors of Year 2 observations, including 12 sectors at 2-minute cadence (14--24 and 26), and all sectors in the 30-minute cadence Full Frame Images (FFIs). In Year 4 of TESS operations, TOI-2095 was observed for 11 sectors (40, 41, and 47--55) in all cadences (20-second short cadence, 2-minute short cadence, and 10-minute FFIs). Observations of TOI-2095 continued in Cycle 5 for 6 sectors (Sectors 56-60, observed from August 2022-January 2023), but due to the availability of data during our analyses, these have not been included. 

TOI-2095 was observed at higher cadence (2-min and 20-second cadence observing modes) as a result of its inclusion in a number of TESS Guest Investigator programs including G022198 (PI: Dressing, 2-min only) in Cycle 2 and G04006 (PI: Ramsay), G04039 (PI: Davenport), G04129 (PI: Buzasi), G04148 (PI: Robertson), G04178 (PI: Pepper), G04191 (PI: Burt), and G04242 (PI: Mayo) in Cycle 4. Both TOIs (TOI-2095.01 and TOI-2095.02) were alerted on 2020 July 15 by TESS \citep{Guerrero2021} following detection by the TESS pipeline \citep{Jenkins2016}. It was at this time that we began investigating this planetary system further.

\begin{table}[h]
  \centering
  \caption{TOI-2095 lies in the Northern Continuous Viewing Zone and was observed by TESS in 29 sectors beginning 2019-Jul-18. The analysis presented herein includes data from Cycle 2 to Cycle 4.}
  \begin{tabular}{cccc |ccccccc}
    Cycle & Sector & Camera & Cadence(s) & & & & Cycle & Sector & Camera & Cadence(s)\\
    \hline
     2 & 14 & 3 & 2min/FFI & & & & 4 & 47 & 4 & 2min/FFI/20s\\
     2 & 15 & 3 & 2min/FFI & & & & 4 & 48 & 4 & 2min/FFI/20s \\
     2 & 16 & 3 & 2min/FFI & & & & 4 & 49 & 4 & 2min/FFI/20s \\
     2 & 17 & 4 & 2min/FFI & & & & 4 & 50 & 4 & 2min/FFI/20s \\
     2 & 18 & 4 & 2min/FFI & & & & 4 & 51 & 4 & 2min/FFI/20s \\
     2 & 19 & 4 & 2min/FFI & & & & 4 & 52 & 3 & 2min/FFI/20s \\
     2 & 20 & 4 & 2min/FFI & & & & 4 & 53 & 3 & 2min/FFI/20s \\
     2 & 21 & 4 & 2min/FFI & & & & 4 & 54 & 4 & 2min/FFI/20s \\
     2 & 22 & 4 & 2min/FFI & & & & 4 & 55 & 4 & 2min/FFI/20s \\
     2 & 23 & 4 & 2min/FFI & & & & 5 & 56 & 4 & --- \\
     2 & 24 & 3 & 2min/FFI & & & & 5 & 57 & 4 & --- \\
     2 & 25 & 3 & FFI   & & & & 5 & 58 & 4 & --- \\ 
     2 & 26 & 3 & 2min/FFI & & & & 5 & 59 & 4 & --- \\
     4 & 40 & 3 & 2min/FFI/20s & & & & 5 & 60 & 4 & --- \\
     4 & 41 & 3 & 2min/FFI/20s & & & & & & & \\
  \end{tabular}
  \label{tab:sectors}
\end{table}
\vspace{1em}
\section{Stellar Characterization} \label{sec:star}

TOI-2095 is a low-mass star in the constellation of Draco. It was identified as a cool dwarf in the TESS Input Catalog \citep[TIC; ][]{Stassun2019_TICv8,Muirhead2018}. We collated archival data on this source and combined the TIC information with newly collected data to derive stellar properties using multiple different approaches. 

We used \textit{Gaia} astrometric and radial velocity (RV) data from Data Release 3 \citep[DR3,][]{TheGaiaMission2016, GaiaCollab2022}. The astrometry and RV values were combined to determine the Galactic kinematics, presented in Table~\ref{table:hoststarinfo_observed}.

Photometric brightness values reported in Table~\ref{table:hoststarinfo_observed} were drawn from \textit{Gaia} DR3, the TIC, the Two Micron All-Sky Survey \citep[2MASS,][]{Cutri2003_2MASS,Skrutskie2006_2MASS}, and the \textit{Wide-field Infrared Survey Explorer} (WISE) AllWISE data release \citep{Wright2010,Cutri2014_AllWISE_VizieR}. Using the \citet{PecautMamajek2013_stellar_properties_table} color-temperature table\footnote{\scriptsize \url{http://www.pas.rochester.edu/~emamajek/EEM_dwarf_UBVIJHK_colors_Teff.txt}} and the available photometry, we estimated a spectral type of M1V.

\begin{deluxetable*}{lccl}
\tabletypesize{\small}
\tablecaption{Host Star Observed Properties and Literature Values}
\setlength{\tabcolsep}{0.03in}
\tablewidth{0pt}
\tablehead{
\colhead{Property} &
\colhead{Value} &
\colhead{Error} &
\colhead{Ref.}}
\startdata
\hline
\hline
Identifiers & \multicolumn{3}{l}{TOI-2095, TIC~235678745, \textit{Gaia}~DR2~2268372099615724288} \\
\hline
\hline
\multicolumn{4}{c}{Astrometry and Kinematics}\\
\hline
\hline
R.A. J2016 (deg)  & 285.63663592007 & 0.0000000036 & \textit{Gaia} DR3  \\% R.A. J2015.5 (deg)  & 285.63652352664 & 0.0292  & \textit{Gaia} DR2  \\ % 
Decl. J2016 (deg)  & +75.41851051257 & 0.0000000035 & \textit{Gaia} DR3  \\ %Decl. J2015.5 (deg) & +75.41851353400 & 0.0270  & \textit{Gaia} DR2  \\ % 
Parallax (mas) & 23.8571 & 0.0133 & \textit{Gaia} DR3 \\%23.8274  & 0.0282  & \textit{Gaia} DR2  \\ % 
Distance (pc) & 41.917 & 0.023& \textit{Gaia} DR3 \\
R.A. Proper Motion (mas yr$^{-1}$) & 203.466 & 0.018 & \textit{Gaia} DR3 \\%203.440 & 0.048 & \textit{Gaia} DR2  \\ %
Decl. Proper Motion (mas yr$^{-1}$) & -21.401 & 0.017 & \textit{Gaia} DR3 \\%-21.318 & 0.047 & \textit{Gaia} DR2  \\ %	  % Total Proper Motion (mas yr$^{-1}$) & 204.588 -- & \textit{Gaia} EDR3 % no PMtot error or position angle
% Tangential Velocity (km s$^{-1}$) & 40.692 & 0.049 & This work (using \textit{Gaia} DR2 values) \\ % Can be removed along with text once RV and UVW are added
Radial Velocity (km s$^{-1}$) & -19.94 & 0.52 & \textit{Gaia} DR3 \\ %\nodata & \nodata & \nodata \\ %
U$_{LSR}$ (km s$^{-1}$) &   5.95 & 0.32 & \textit{This work} \\
V$_{LSR}$ (km s$^{-1}$) &  12.05 & 0.62 & \textit{This work} \\
W$_{LSR}$ (km s$^{-1}$) & -38.70 & 0.34 & \textit{This work} \\
Total Galactic Motion (km s$^{-1}$) & 40.97 & 0.37 & \textit{This work} \\
\hline
\hline
\multicolumn{4}{c}{Photometry}\\
\hline
\hline
$BP$ & 12.086019 & 0.002779 &\textit{Gaia} DR3 \\ %$BP$ & 13.1291 & 0.0021 & \textit{Gaia} DR2  \\ % 
$RP$ & 13.111851 & 0.003183 & \textit{Gaia} DR3 \\ %$RP$ & 11.1002 & 0.0013 & \textit{Gaia} DR2  \\ %
$G$  & 11.080335 & 0.003842 & \textit{Gaia} DR3 \\ %$G$  & 12.0940 & 0.0004 & \textit{Gaia} DR2   \\ 
$V_J$ & 12.854 & 0.030 & \textit{Gaia} DR3 Conversion \\%$V_J$    & 12.838 & 0.046 & \textit{Gaia} DR2 Conversion \\ 
$R_{KC}$ & 11.870 & 0.032 & \textit{Gaia} DR3 Conversion \\ %$R_{KC}$ & 11.873 & 0.048 & \textit{Gaia} DR2 Conversion \\ 
$I_{KC}$ & 10.934 & 0.038 & \textit{Gaia} DR3 Conversion \\ %$I_{KC}$ & 10.964 & 0.050 & \textit{Gaia} DR2 Conversion \\ 
$T$ & 11.0788 & 0.0073 & TIC v8.1 \\
$J$ & 9.797 & 0.020 & 2MASS\\
$H$ & 9.186 & 0.015 & 2MASS\\
$K_s$ & 8.988 & 0.015 & 2MASS \\
$W1$ & 8.868 & 0.025 & AllWISE \\
$W2$ & 8.766 & 0.024 & AllWISE \\
$W3$ & 8.670 & 0.024 & AllWISE \\
$W4$ & 8.709 & 0.269 & AllWISE \\		
\hline
\hline
\multicolumn{4}{c}{Spectral Features}\\
\hline
\hline
Spectral Type & M0.5V-M1.0V & \nodata & \citet{PecautMamajek2013_stellar_properties_table} Table$^*$ \\
\hline
\multirow{2}{*}{Metallicity ([Fe/H]; dex)} & subsolar & \nodata & This work: HR Diagram, SED Fits, etc. \\
 %& -0.48 & 0.20 & \cite{Bonfils2005_MetaPhotRelation} relation \\
% & -0.19 & 0.06 & \citet{Mann2013a} relation \\
 & -0.45 & 0.38 & \cite{Kesseli2019} relation \\
\hline
\enddata
\tablenotetext{}{2MASS: \cite{Cutri2003_2MASS,Skrutskie2006_2MASS}, WISE AllWISE Release: \cite{Wright2010,Cutri2014_AllWISE_VizieR}}
\tablenotetext{*}{\href{http://www.pas.rochester.edu/~emamajek/EEM_dwarf_UBVIJHK_colors_Teff.txt}{http://www.pas.rochester.edu/$\sim$emamajek/EEM\_dwarf\_UBVIJHK\_colors\_Teff.txt}}
\label{table:hoststarinfo_observed}
\end{deluxetable*}

We use multiple different approaches to estimate stellar parameters to allow cross-comparison and identify any outliers from the choice of parameter estimation method. Mass, radius, effective temperature, and luminosity are calculated using relations presented in \cite{Mann2015, Mann2019}, \cite{Benedict2016}, and an expanded version of the method presented in \cite{Dieterich2014} from \cite{Silverstein2019}. We estimate the stellar metallicity using the relations from \cite{Kesseli2019}. When the parameter estimation approach requires absolute magnitudes, we use the most recent TOI-2095 parallax from Gaia DR3 \citep{TheGaiaMission2016, GaiaCollab2022}

Following the mass-$M_K$ relation of \cite{Mann2019}, we derived a mass of 0.465~$\pm$~0.012~$M_\odot$. For this calculation, we computed $M_K$ using the 2MASS $K_s$ magnitude and \textit{Gaia}~DR3 parallax, and adopted the \cite{Kesseli2019} metallicity, [Fe/H] = -0.45$\pm$0.38 dex. We found only a 0.001~$M_\odot$ difference in the mass result when [Fe/H]=0~dex was adopted. For comparison, we also calculate masses using the \cite{Benedict2016} mass-$M_K$ and mass-$M_V$ relations, which do not include a correction for metallicity. In agreement with the systematic comparison of the two relations in \cite{Mann2019}, the $M_K$ mass value of $0.518\pm0.020~M_\odot$ from the \cite{Benedict2016} relation is higher than that from the \cite{Mann2019} relation. \cite{Mann2019} notes that the difference in the two mass-$M_K$ relations likely stems from improvements in the observations of stars with masses $\gtrsim0.35~M_\odot$ used to calibrate each method. We also derive a \cite{Benedict2016} mass-$M_V$ relation value of $0.489\pm0.021~M_\odot$, with $M_V$ calculated using a \textit{Gaia}~DR3 parallax and a $V$ magnitude calculated using the \textit{Gaia}~DR3 photometry and conversions via the \cite{Riello2021_GaiaEDR3_PhotConversions} relations. This value matches the \cite{Benedict2016} mass-$M_K$ result within the uncertainties.

We use the $M_K$ relations in \cite{Mann2015} to estimate luminosity, effective temperature and radius. These calculations use a semi-empirical approach to parameter estimation calibrated with a sample of M dwarfs having interferometric radius measurements and also take into account the \cite{Kesseli2019} metallicity. The estimated values are provided in Table~\ref{table:hoststarinfo_derived}.

We also derived effective temperature, luminosity, and radius following the method described in \cite{Silverstein2019}, which is similar to that described in \cite{Dieterich2014}. We compared observed $VRI$, 2MASS $JHK_s$, and WISE $W1W2W3$ photometry to photometry extracted from the BT-Settl 2011 photospheric models \citep{Allard2012} to determine an effective temperature. Observational $VRI$ were derived according to the \textit{Gaia}~DR2 photometry conversion. The closest-matching model spectrum to our T$_{\rm eff}$ was iteratively scaled by a polynomial until observations match the scaled model. The spectrum was then integrated within the wavelength range of the observations. A bolometric correction was applied based on how much blackbody flux would be missing beyond the observed wavelength range for a star of the same T$_{\rm eff}$. Bolometric flux was then scaled by the \textit{Gaia}~DR3 parallax to produce bolometric luminosity. Lastly, effective temperature and luminosity were substituted into the Stefan-Boltzmann law to calculate the stellar radius.

Using this method, we derived two sets of values for different adopted model metallicities, [Fe/H]=0~dex and [Fe/H]=-0.5~dex. We compared these results to those derived using the \cite{Mann2015} relations, which also include a metallicity term. We found that the results were a better match when both methods adopted a [Fe/H] value of -0.5 dex, with T$_{\rm eff}$ and radius values nearly identical. Derived stellar properties are reported in Table~\ref{table:hoststarinfo_derived}. We adopt the \cite{Mann2015, Mann2019} values for subsequent analyses in this paper due to their more conservative uncertainties which take into account the scatter in the calibration samples used in those relations. 

\begin{deluxetable*}{lccc}%[!h]
\tabletypesize{\small}
\tablecaption{Host Star Derived Properties (with adopted values in bold)}
\setlength{\tabcolsep}{0.03in}
\tablewidth{0pt}
\tablehead{
\colhead{Property} &
\colhead{Value} &
\colhead{Error} &
\colhead{Ref.}}
\startdata
\hline
Names & \multicolumn{3}{l}{TOI-2095, TIC~235678745, \textit{Gaia}~DR2~2268372099615724288} \\ 
\hline
\multirow{3}{*}{Mass ($M_\odot$)} &  \textbf{0.465} & \textbf{0.012} & \cite{Mann2019} Relation ([Fe/H]=-0.45) \\
& 0.518 & 0.020 & \cite{Benedict2016} $\mathcal{M}_K$ Relation \\ %absK=5.873
& 0.489 & 0.021 & \cite{Benedict2016} $\mathcal{M}_V$ Relation$^\dagger$ \\ %absV=9.723
\hline
\multirow{2}{*}{Effective Temperature ($K$)} & \textbf{3662} & \textbf{130} & \citet{Mann2015}~Relation~([Fe/H]~=~-0.45) \\
& 3690 & 50 & \cite{Silverstein2019} ([Fe/H]~=~-0.5) \\
\hline
\multirow{2}{*}{Luminosity ($L_\odot$)} & \textbf{0.0333} & \textbf{0.0019} & \citet{Mann2015}~Relation~([Fe/H]~=~-0.45) \\
& 0.0345 & 0.0008 & \cite{Silverstein2019} ([Fe/H]~=~-0.5) \\
\hline
\multirow{2}{*}{Radius ($R_\odot$)} & \textbf{0.453} & \textbf{0.029} & \citet{Mann2015}~Relation~([Fe/H]~=~-0.45) \\
& 0.454 & 0.014 & \cite{Silverstein2019} ([Fe/H]~=~-0.5) \\
\hline
Density (g~cm$^{-3}$) & \textbf{7.05} & \textbf{1.38} & Using Mann~et al. M \& R \\
\enddata
%\tablenotetext{}{adopted values in bold}
\tablenotetext{\dagger}{Using a $V$ magnitude converted from \textit{Gaia}~DR3 photometry.}
\label{table:hoststarinfo_derived}
\end{deluxetable*}

TOI-2095 is likely slightly older and metal-poor compared to other M dwarfs of its spectral type. When plotted using \textit{Gaia} DR3 photometry and astrometry, the star is relatively low on the color-magnitude diagram, common for lower metallicity stars. A lack of starspot-induced photometric modulation and flares indicates low magnetic activity, more commonly seen in older stars. As described previously, agreement between methods to derive fundamental properties improves when subsolar metallicity is adopted. Photometric metallicities also suggest that the star is slightly metal-poor  (Table~\ref{table:hoststarinfo_observed}). The total Galactic space motion of $\sim$41~km~s$^{-1}$ does not suggest that the star is a member of the thick disk or Galactic halo.
\vspace{2em}
\section{Ground Based Observations for System Validation} \label{sec:followup}

In addition to observations from TESS, we collected numerous additional datasets in support of determining whether the two planetary transit signals are from bona fide planets. These data comprise high contrast imaging, spectroscopy of the star, and photometry of the star collected during transits.

\subsection{Imaging}

\begin{figure}
    \centering
    \includegraphics[width=\textwidth]{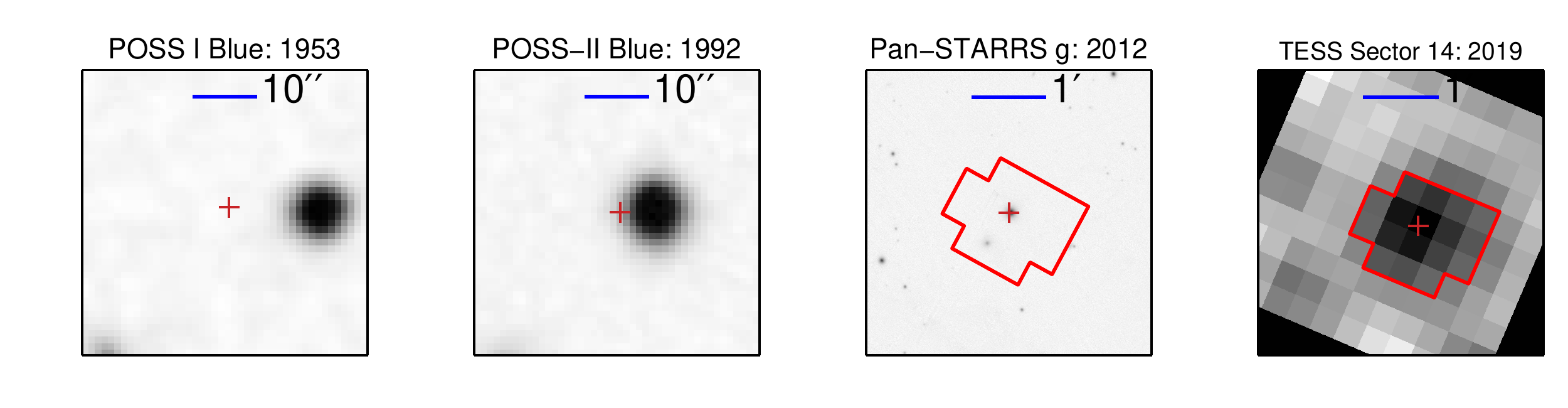}
    \caption{Archival imaging of TOI 2095. From left to right, an image from the first Palomar Observatory Sky Survey (POSS-I) in a blue-sensitive emulsion, an image from the second Palomar Observatory Sky Survey (POSS-II) in a blue-sensitive emulsion, an image from the PAN-STARRS telescope in $g$-band, and the summed image from TESS during Sector 14. The present-day position of TOI 2095 is shown with a red cross, and in the Pan-STARRS and TESS images, the outline of the TESS photometric aperture from Sector 14 is shown as a solid red line. TOI 2095's high proper motion reveals no background sources at its present-day position that could give rise to the transits we see.}
    \label{fig:possimaging}
\end{figure}

TOI-2095 has large proper motion (205.6 mas/yr) and has moved significantly since the first archival sky survey observations in the 1950s (see Figure~\ref{fig:possimaging}). Within the $\sim$5 pixel TESS aperture ($\sim$105 arcsec box) there are no background stars at the current location of TOI-2095 down to the background limits of the POSS-I surveys ($\sim$20 mag in B and R). There are 2 faint background galaxies within the aperture, but these cannot contribute as source of false positives for the TESS detected transits.

High-contrast imaging enables us to infer limits on the brightness and separations of any bound, co-moving companions. To do so, we collected data using adaptive optics (AO) and speckle imaging techniques.

\subsubsection{AO Imaging}

 We used the NIRC2 instrument behind the Natural Guide Star (NGS) AO system \citep{Wizinowich2000} on the 10 m Keck-II telescope. We obtained data on 2020 September 09 UT in the $K$-band filter with $\lambda_c$~= 2.196~$\mu$m and bandwidth 0.336 $\mu$m under clear skies. We followed the general observation plan and analysis approach described in \cite{2021FrASS...8...63S} for NIRC2 high resolution imaging of TESS systems. Briefly, we observed using 0.181 second integrations following a standard dither sequence comprised of 3 arcsec steps that were repeated three times, with each subsequent dither offset 0.5 arcsec. At each location we used 1 co-add, resulting in 9 total frames. We used the narrow-angle mode of the NIRC2 camera which has a plate scale of 9.942 milliarcseconds pixel$^{-1}$ and a 10 arcsec FOV. No companions were detected down to a contrast of 6 magnitudes at 0.2'' (8.4 AU separation at the distance of TOI-2095).

\begin{figure}
	\centering
	\includegraphics[width=0.5\textwidth]{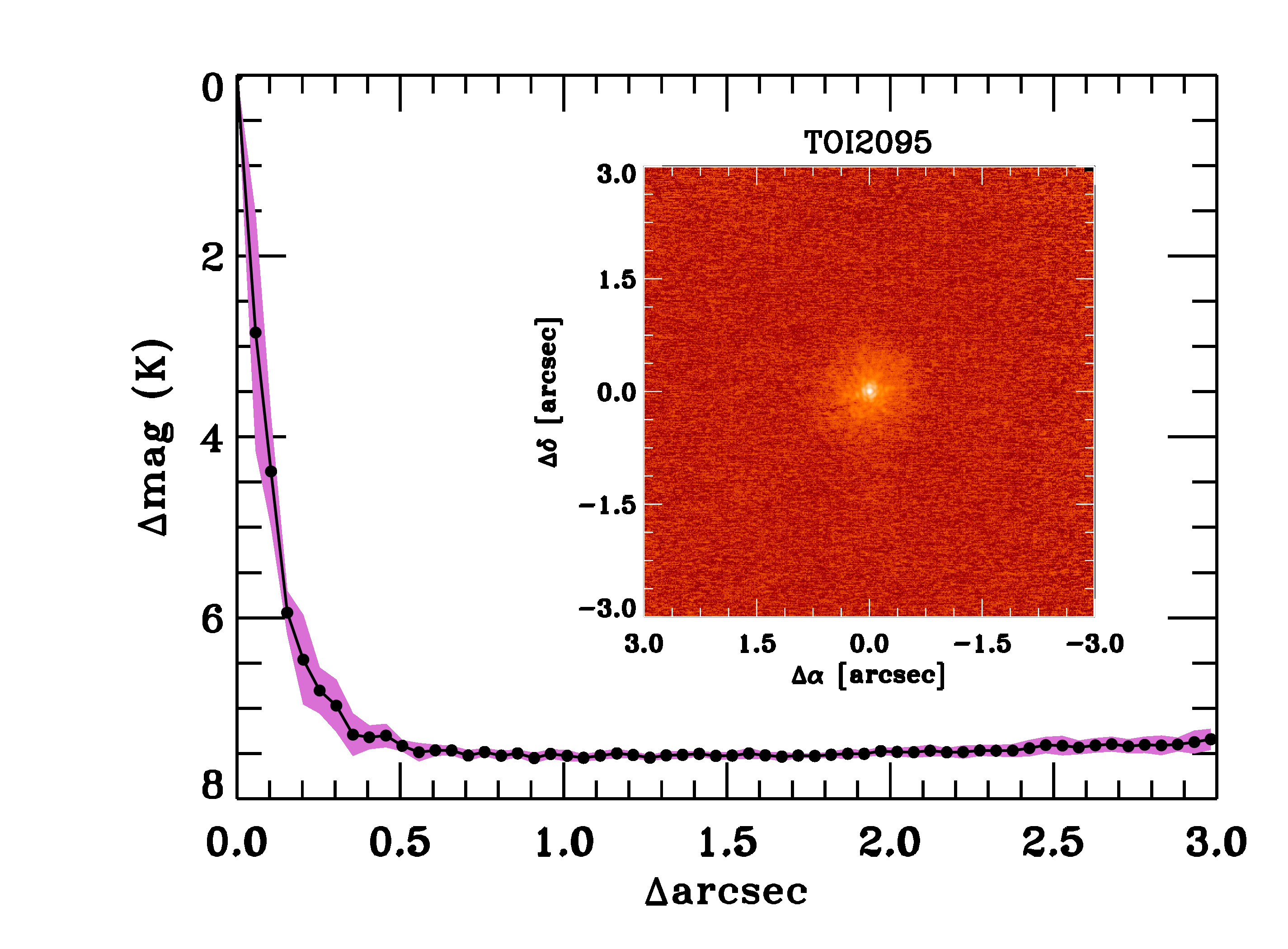}
	\caption{Keck-II NIRC2 $K$-band adaptive optics image of TOI-2095 (inset) and contrast curve. No faint companions were detected down to 6 mag of contrast at separations of 0.2$^{\prime\prime}$ and 7.5 mag at separations $>$0.5$^{\prime\prime}$}
	\label{fig:keck}
\end{figure}

\subsection{Speckle imaging}

\begin{figure}
	\centering
	\includegraphics[width=0.5\textwidth]{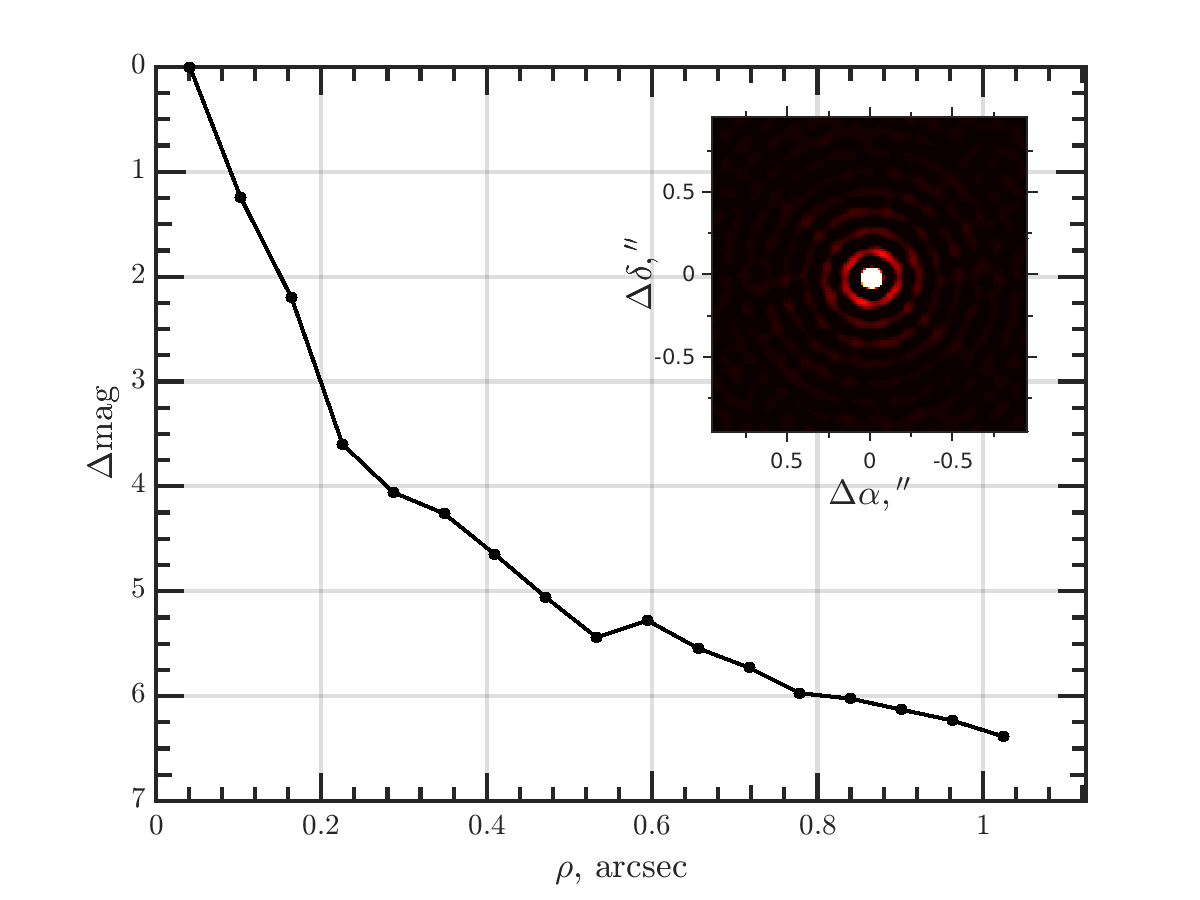}
	\caption{Speckle SAI image of TOI-2095 (inset) and contrast curve.}
	\label{fig:sai}
\end{figure}

TOI-2095 was observed on 2020 December 26 UT with the Speckle Polarimeter \citep{2017AstL...43..344S} on the 2.5~m telescope at the Caucasian Observatory of Sternberg Astronomical Institute (SAI) of Lomonosov Moscow State University. SPP uses Electron Multiplying CCD Andor iXon 897 as a detector. The atmospheric dispersion compensator allowed observation of this relatively faint target through the wide-band $I_c$ filter. The power spectrum was estimated from 4000 frames with 30 ms exposure. The detector has a pixel scale of $20.6$ mas pixel$^{-1}$, and the angular resolution was 89 mas. We did not detect any stellar companions brighter than $\Delta I_C=3.8$ and $6.4$ at $\rho=0\farcs25$ and $1\farcs0$, respectively, where $\rho$ is the separation between the source and the potential companion, see Figure~\ref{fig:sai}.

\subsection{Additional Photometry}

The \textit{TESS} pixel scale is $\sim 21\arcsec$ pixel$^{-1}$ and photometric apertures typically extend out to roughly 1 arcminute, generally causing multiple stars to blend in the \textit{TESS} aperture. To attempt to determine the true source of the detections in the \textit{TESS} data and refine their ephemerides and transit shapes, we conducted ground-based photometric follow-up observations of the field around TOI-2095 as part of the {\tt TESS} Follow-up Observing Program\footnote{https://tess.mit.edu/followup} Sub Group 1 \citep[TFOP;][]{collins:2019}. We used the {\tt TESS Transit Finder}, which is a customized version of the {\tt Tapir} software package \citep{Jensen:2013}, to schedule our transit observations. Differential photometric data were extracted using {\tt AstroImageJ} \citep{Collins:2017}.

Observations were collected during a transit of TOI-2095 b on 2020 September 23 UT from Campo Catino Rodeo Observatory in Rodeo, New Mexico using a remotely operated Planewave 35 cm telescope. The observations used a clear filter and 180-second exposures. This observation consists of seeing limited photometry of the system and the surrounding stars in order to check for nearby eclipsing binaries (NEB), following the standard TESS Follow-up Program (TFOP) Sub-Group 1 (SG1) procedure.

The target lightcurve was flat with with a scatter of 0.77 ppt with a 530 second cadence using a 13px=5.3 arcsec aperture. This was close to the predicted depth (0.79 ppt), but no significant transit ingress was visible. A NEB check was performed using a 5px=2.5 arcsec aperture and did not find any obvious NEBs.

A second epoch of NEB checking for TOI-2095 b was performed with the Las Cumbres Observatory Global Telescope \citep[\lco;][]{Brown:2013} 0.4-meter telescope at Haleakala on 2021 May 29. No eclipsing binaries were identified within 2.5 arcmin of the target.

Two observations were also collected with the 1.0-meter \lco\ telescope of McDonald Observatory using the the Sinistro instrument with the SDSS i$^\prime$ filter. The first of these covered a full transit of planet c on 2020 September 06 UT, and the second covered a partial transit of planet b on 2020 September 24 UT. %A transit is visible by eye in these two datasets. 
 These transits are used in the light curve model in Section~\ref{sec:transit}. These data confirm that the transit events occur on-target relative to all known Gaia DR3 and TICv8 stars. The photometric aperture sizes used to extract the on-target detections were 7.0 arcsec for the 2020-09-06 observations and 4.7 arcsec for the 2020-09-04 observations. 

On 2021 February 22 UT a partial transit of TOI-2095 c was observed from the 1.6-m telescope at Observatoire du Mont-M\'egantic (OMM). The photometric aperture radius was 10 pixels (4.66''). A partial transit is visible but the observations were ended prior to the completion of the transit due to high humidity. This partial transit was included in the transit timing variations analysis described in Section~\ref{sec:ttvs}.

% \subsection{Reconnaissance Spectroscopy}%
% \label{ssec:spectroscopy}
% On 2020 September 25, we obtained a TRES spectrum with the FLWO (1.5m). 

\section{Vetting and Validation}
%\section{Statistical Vetting with DAVE and \texttt{Vespa}} 

\begin{figure*}
  \centering
  \includegraphics[width=0.3\textwidth]{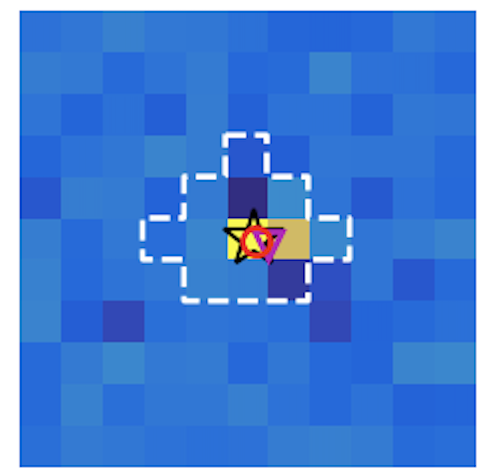}
  \includegraphics[width=0.45\textwidth]{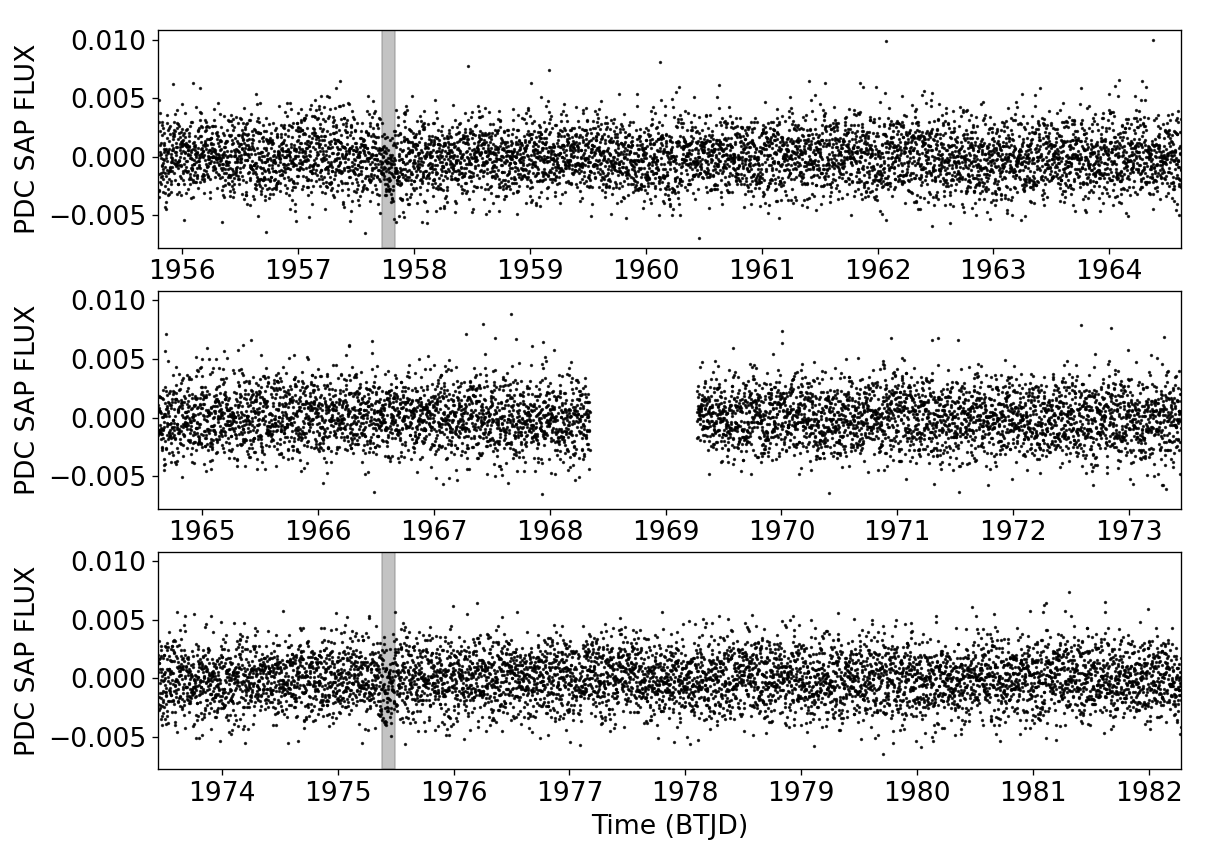}
  \includegraphics[width=0.4\textwidth]{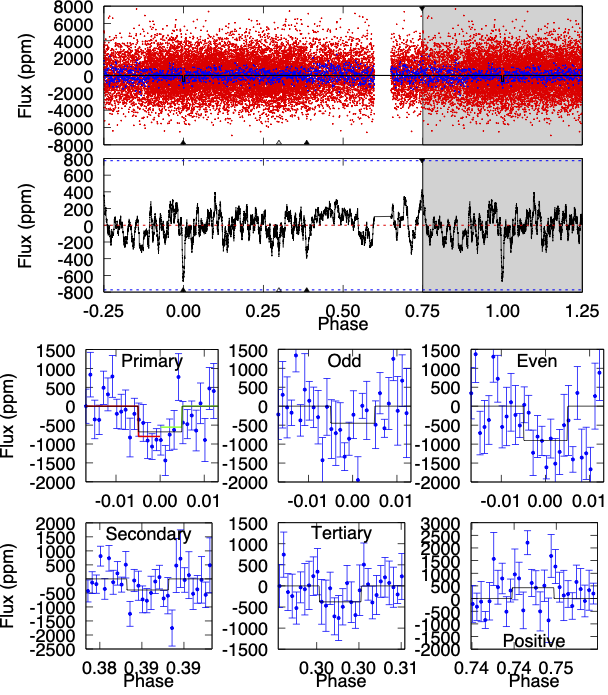}
  \includegraphics[width=0.4\textwidth]{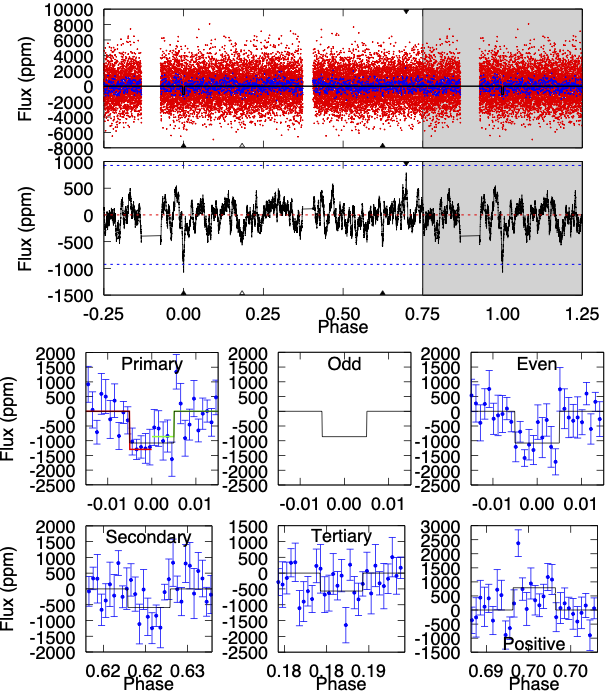}
  \caption{There are only two transits for TOI 2095.01 and only one for TOI 2095.02 in this sector, all with fairly low SNR, but there are no indications for false positives. Left: The difference image for TOI-2095 planet 1 for Sector 24, showing the position of the target (black star), the measured difference image photocenters from a point-spread-function and pixel-response-function (cyan triangle and red circle, respectively). Right: PDC SAP lightcurve of TOI-2059 for Sector 24 highlighting the two transits in the sector. Lower panels: Modshift results from DAVE, left for TOI-2095.01 and right for TOI 2095.02, respectively. The small panels show zoom-ins and corresponding fits to the primary transits, the odd and even transits, the most significant secondary transits, as well as to any additional negative (tertiary) or positive events.} 
  \label{fig:modshift_01}
\end{figure*}

%\comment{This TOI was in Luca's paper}
Due to the shallow transits and long orbital periods of the two planets, the centroid and modshift measurements from the Discovery and Vetting of Exoplanets (DAVE) vetting pipeline \citep{Kostov2019} are somewhat unreliable. This is demonstrated in Figure \ref{fig:modshift_01} showing the results for planet b for Sector 24. While the data are low SNR and only have two transits in this sector (shown in the upper right panel), there are no indications for a false positive. 

A false alarm detection due to random noise is highly unlikely, as both planets were detected in ground-based transits. However, a scenario where the signals detected are caused by background contaminating sources always exists. Having signals from two planets, as we have here, would require either two background eclipsing binaries or a background system of planets, both of which are unlikely scenarios. To demonstrate this we look at the statistical probability that the detections are false positives.

We statistically analyzed the likelihood of false-positive signals using the publicly-available software package \texttt{vespa} \citep{2012ApJ...761....6M, 2015ascl.soft03011M} to place a numerical value on the false positive probability (FPP) of the signals. \texttt{vespa} combines the host star properties, the observed TESS transits, and follow-up constraints to compare the signals to six astrophysical false positive scenarios allowed by the remaining parameter space in a probabilistic framework: an unblended eclipsing binary (EB), a blended background EB, a hierarchical EB companion, and double period scenarios of these three. The output of \texttt{vespa} is the likelihood that a detected transit signal may be mimicked by one of these astrophysical false positive scenarios. \texttt{vespa} assumes that the signal is coming from the target star, which was shown to be true for both signals in Section \ref{sec:followup}. The software was run on each transit signal individually after masking out transits from other planets. We included observational constraints in our analysis along with the addition of the Keck contrast curve (see Section \ref{sec:followup}). % as well as radial velocity constraints derived from the TRES data \textcolor{red}{section?}
The final FPP values are 7.94$\times$10$^{-5}$ and 4.87$\times$10$^{-4}$ for TOI-2095 b and TOI-2095 c, respectively. Both of these are $\ll$0.01 and thus constitute firm validations of the planetary scenario.

As an independent check, we also ran these signals through another statistical validation software, TRICERATOPS \citep{giacalone2020vetting}. Developed specifically for use with TESS data, TRICERATOPS differs from \texttt{vespa} in that it accounts for the TESS extraction aperture and the actual background starfield when calculating the likelihood that the signal originates from a star nearby in the field of view. Otherwise, TRICERATOPS works similarly, testing the shape and depth of the transit against a suite of possible astrophysical false positive scenarios to provide a final FPP as well as a Nearby FPP (NFFP) which is the probability that the signal is due to a false positive scenario around a nearby star. We ran TRICERATOPS using the same inputs as VESPA with the addition of the apertures used to extract the PDCSAP light curves generated by the TESS SPOC. We find that TOI-2095 b has an FPP of 1.85$\pm$1.11 $\times$10$^{-3}$ and an NFPP of 1$\pm$1 $\times$10$^{-8}$ while for TOI-2095 c has an FPP of 7.17$\pm$7.03 $\times$10$^{-2}$ and an NFPP of $<$1$\times$10$^{-8}$. Since TOI-2095 b has an FPP $<$ 0.015 and an NFPP $<$ 10$^{-3}$, this signal is considered validated. However, because the FPP for TOI-2095 c is greater than the 0.015 threshold but is still less than 0.5 with an NFPP $<$ 10$^{-3}$, this signal is only a likely planet. 

However, neither the FPPs from \texttt{vespa} nor those from TRICERATOPs account for the fact that this system is host to multiple signals, implying a lower FPP by $\sim$50$\times$ due to what is termed a ``multiplicity boost" \citep{Lissauer2012,Guerrero2021}. Since this puts the FPP values for both planet candidates $\ll$1$\%$, we consider these signals to be validated planets.

\section{Data Analysis and Results}

\subsection{Light Curve Model}
\label{sec:transit}
We modeled the transits of the two planets in the TOI-2095 system using a Bayesian framework to compute stellar and exoplanet parameters from a limb darkened light curve model \citep{exoplanet:luger18,exoplanet:agol20}. We assumed a linear ephemeris, and used initial parameters computed by the TESS pipeline \citep{Jenkins2016}, combined with the stellar properties calculated in Section~\ref{sec:star}.

Our transit model uses the TESS 2-minute cadence PDC SAP data \citep{Stumpe2012, Stumpe2014, Smith2012}, plus two observations from \lco. For a linear ephemeris, the addition of the available 20-s cadence data does not provide a significant improvement in the model-fit. However, for the later analysis of transit timing variations (TTVs) in Section~\ref{sec:ttvs} we do include the higher cadence data. We also do not include the partial transit of planet c from OMM here, although we again include it in the TTV section.

The parameters included in the model are: the stellar radius and density, two stellar limb darkening parameters for each instrument (i.e for TESS and \lco), the photometric zeropoint (one for the entire TESS dataset and one for each ground-based observation), and terms for additional white noise for each separate instrument configuration. In addition, for each planet we include a transit midpoint, orbital period, impact parameter, planet radius in units of the stellar radius, and two eccentricity vectors (e$\sin\omega$ and e$\cos\omega$ ). We use a Gaussian Process (GP) to model any variability in the time series that is not described by the model. The GP is a kernel stochastically-driven, damped harmonic oscillator using the \texttt{celerite} package \citep{exoplanet:foremanmackey17,exoplanet:foremanmackey18}, similar to the model described in \citep{Gilbert2020}. Separate hyperparameters are used for the GPs applied to TESS and \lco\ data.

The priors on the model parameters are: Gaussian for stellar radius, orbital period, mid-transit time, and photometric zeropoint; Log-normal for the stellar density and scaled planet radius; Uniform between zero and one for the impact parameter; the limb darkening parameters follow \citet{Kipping2013}. We include two components in a prior on eccentricity: we have a 1/eccentricity prior owing to the bias of sampling in vector space, and we include a Beta prior following \citet{exoplanet:kipping13} with hyperparameters from \citet{exoplanet:vaneylen19} using the values for multiplanet systems.

We built this model in the \texttt{exoplanet} software \citep{exoplanet:exoplanet} which is built on \texttt{PyMC} \citep{exoplanet:pymc3}, a Python library that allows users to build Bayesian models and sample them uses Markov Chain Monte Carlo methods. We use the No U-turn Sampler \citep[NUTS;][]{NUTS} which is a form of Hamiltonian Monte Carlo. We used 3000 samples to tune the posterior and 3000 to sample the posterior distribution. We did this for four independent chains, so that we could use these to test for convergence. All chains had consistent results, so we combined the chains into a single set of posterior samples.

\begin{figure}
    \centering
    \includegraphics[width=0.5\textwidth]{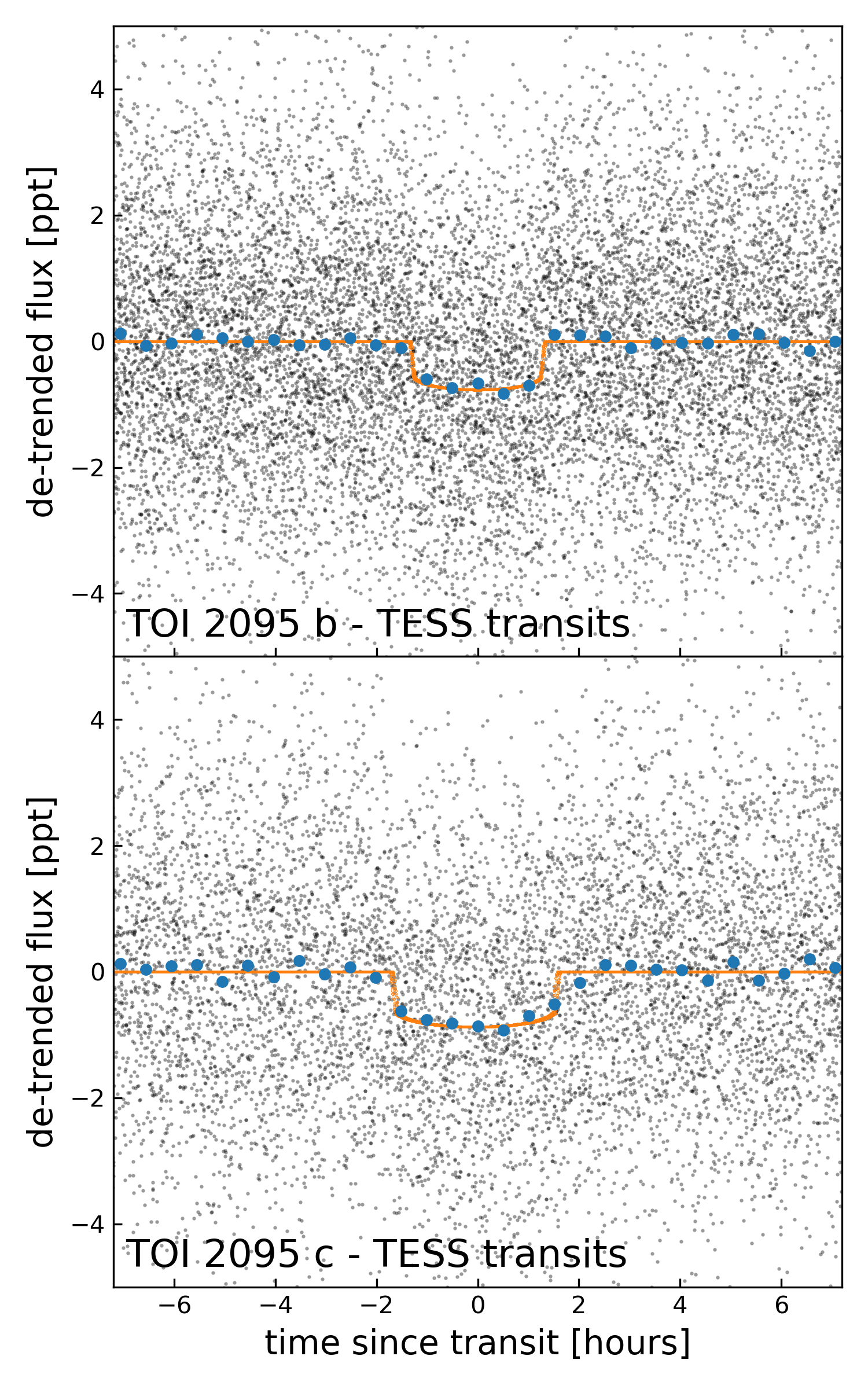}
    \caption{Folded TOI-2095 b and c transits observed by TESS. The grey points are the observed data, folded on the best-fitting orbital period, the blue dots are binned data, and the orange curve is the best fitting transit model.}
    \label{fig:TESS-transits}
\end{figure}

\begin{figure}
    \centering
    \includegraphics[width=0.5\textwidth]{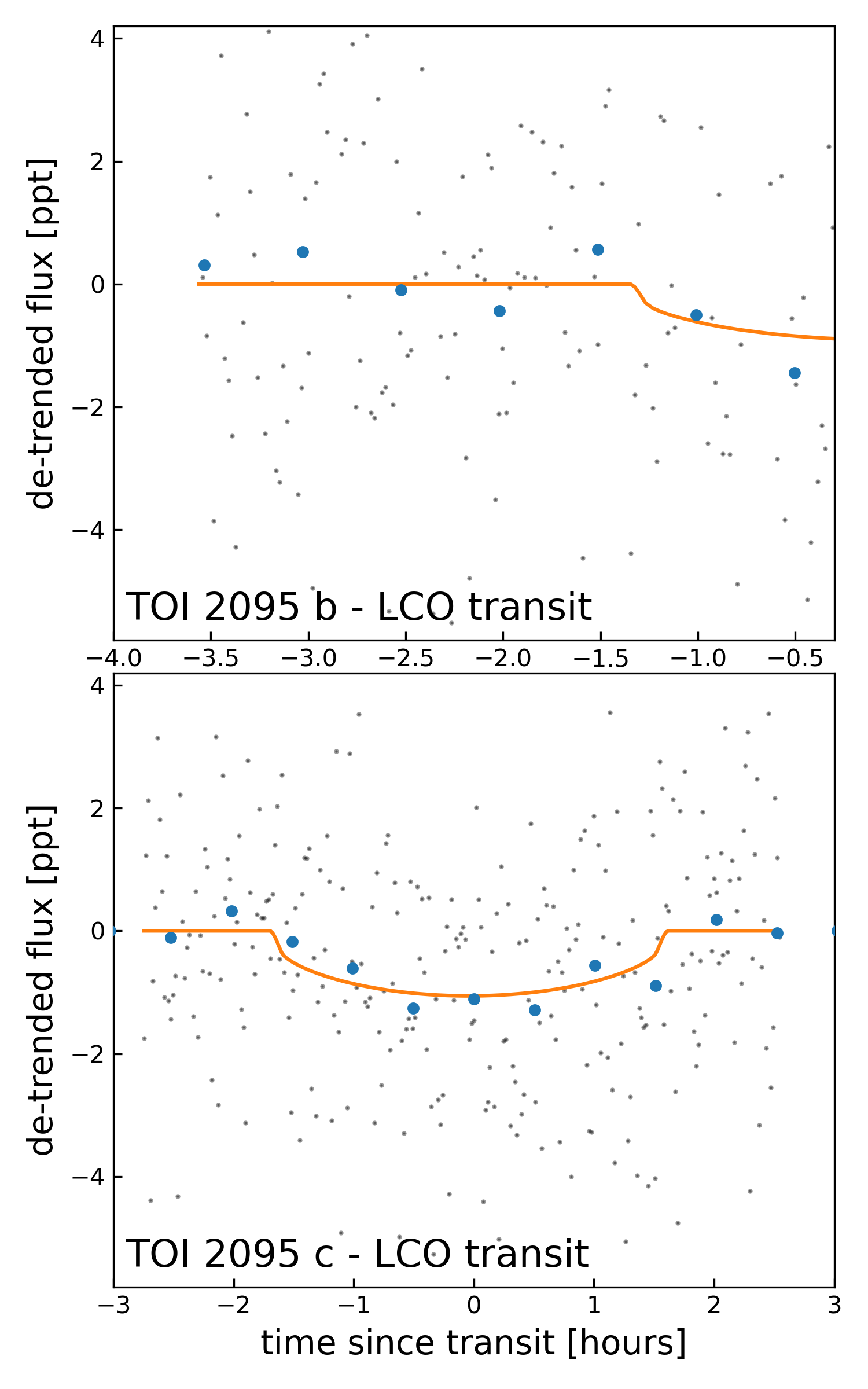}
    \caption{Two transit of TOI-2095 observed by \lco. The upper panel shows the partial transit of planet b and the lower panel shows the full transit of planet c.}
    \label{fig:LCO-transits}
\end{figure}

The results of this sampling are provided in Table~\ref{tab:mcmc}. The two planets have radii of $1.30\pm 0.10$ and $1.39\pm 0.11$ R$_\oplus$. Folded transits from the TESS data are shown in Figure~\ref{fig:TESS-transits}, and the two \lco\ epochs of data are shown in Figure ~\ref{fig:LCO-transits}. The orbital eccentricities of both planets are consistent with zero. We used the samples to compute additional planet parameters, that are also listed in the table: the orbital inclination, semimajor axis, insolation flux, and transit duration. The two planets have insolations of 3.2 and 1.7 times the Earth's insolation from the Sun.

We also ran independent fits of the TOI-2095 system using the Cycle 2 lightcurve from TESS with EXOFASTv2 \citep{Eastman2019} and found the fitted parameters to be consistent within 1 $\sigma$ of the results presented here.

\begin{table}[htb]
    \caption{Parameters measured from the light curve model}
    \centering
    \begin{tabular}{llll}
    \textbf{Parameter} & \textbf{Median} & \textbf{+1$\sigma$} & \textbf{-1$\sigma$}\\
    \hline
    \multicolumn{4}{l}{\textbf{Star}} \\
    Stellar Radius & 0.453 & 0.028 & 0.029 \\
    Stellar Density & 6.7 & 1.2 & 1.1 \\
    TESS Limb Darkening u$_1$ & 0.23 & 0.25 & 0.17\\
    TESS Limb Darkening u$_2$ & 0.12 & 0.33 & 0.23\\
    \lco\ Limb Darkening u$_1$ & 0.927745 & 0.525472 & 0.572409\\
    \lco\ Limb Darkening u$_2$ & -0.19 & 0.52& 0.42\\
    \hline
    \multicolumn{4}{l}{\textbf{TOI-2095 b}} \\
    Transit mid-point (BJD) & 2459646.7038 & 0.0012 & 0.0012\\
    Orbital Period (days) & 17.664872 & 0.000045 & 0.000051\\
    Rp/Rs & 0.0263 & 0.0011 & 0.0011\\
    Impact parameters & 0.30 & 0.22 & 0.20\\
    Eccentricity & 0.12 & 0.19 & 0.08\\
    Argument of periastron (degrees) & -17 & 140 & 130\\
    Radius (R$_\oplus$) & 1.30 & 0.10 & 0.10\\
    a/Rs & 48.0 & 2.7 & 2.7\\
    Semimajor axis & 0.1010 & 0.0088 & 0.0084\\
    Inclination (degrees) & 89.64 & 0.24 & 0.24\\
    Transit Duration (hours) & 2.72 & 0.37 & 0.43\\
    Insolation Flux (S$_\oplus$) & 3.23 & 0.64 & 0.54\\
    \hline
        \multicolumn{4}{l}{\textbf{TOI-2095 c}} \\
    Transit mid-point (BJD) & 2459662.1464 & 0.0018 & 0.0020\\
    Orbital Period (days) & 28.17221 & 0.00011 & 0.00014\\
    Rp/Rs & 0.0282 & 0.0013 & 0.0013\\
    Impact parameters & 0.24 & 0.23 & 0.16\\
    Eccentricity & 0.13 & 0.18 & 0.09\\
    Argument of periastron (degrees) & -55 & 100 & 92\\
    Radius (R$_\oplus$) & 1.39 & 0.11 & 0.10\\
    a/Rs & 65.6 & 3.7 & 3.6\\
    Semimajor axis & 0.138 & 0.012 & 0.011\\
    Inclination (degrees) & 89.79 & 0.14 & 0.18\\
    Transit Duration (hours) & 3.12& 0.39 & 0.54\\
    Insolation Flux (S$_\oplus$) & 1.73 & 0.34 & 0.28\\
    \hline
    \end{tabular}
    \label{tab:mcmc}
\end{table}

\subsection{System Dynamics}\label{sec:dynamics}
We conducted $\sim$10$^{4}$, 1 Myr simulations investigating the long-term dynamical stability of the TOI-2095 system.  Our simulations are based on the $Mercury6$ hybrid integration package \citep{Chambers1999}, and span a range of plausible densities and eccentricities for each planet ($\sim$1-12 $g  cm^{-2}$ and 0.0-0.5, respectively).  This allows us to account for the substantial degeneracy in planet masses \citep[e.g.][]{Chen2017}.  Unique initial conditions for each simulation are created by utilizing each planets' nominal semi-major axis and inclination, and assigning the remaining angular orbital elements randomly by sampling uniform distributions of angles.  

In general, we find the TOI-2095 system to be dynamically stable when the planets' originate on non-crossing orbits ($e\lesssim$0.3 for each planet) for the range of masses we test.  This is not surprising given the well-spaced nature of the system in terms of mutual-Hill radii \citep{chambers96}.  It is important to note that the length of our simulations ($\sim$10$^{7}$ orbits for the inner planet) is likely insufficient to fully characterize system's dynamical state and make a comprehensive determination of its stability \citep[e.g.][]{lithwick11}.  Moreover, we do not consider the possibility of perturbations from other planets in these simulations. Thus, while our simulations do not prove that TOI-2095 is stable, they strongly suggest that it is stable on long timescales.% of $10^7$ orbits.  %Figure \ref{fig:matt_fig} demonstrates an example stable evolution of the system (in this case extended to $t=$ 10 Myr) where the planets begin with $e\simeq$0.10 and Earth-like densities.  While the dynamical evolution is characterized by mild secular oscillations in $e$, it is clear from this example that the system architecture is highly stable on timescales of $10^7$ orbits, even for moderate eccentricities.  

% \begin{figure}
% 	\centering
% 	\includegraphics[width=0.5\textwidth]{figures/matt_fig.pdf}
% 	\caption{Example numerical simulation investigating the dynamical stability of TOI-2095.  In the realization plotted, the planets are assigned Earth-like densities ($\rho \simeq$ 5.5 $g cm^{-2}$) and are initialized on moderately eccentric orbits ($e \simeq $ 0.10).  The perihelion ($q$), semi-major axis and aphelion ($Q$) of each planet is plotted in a different color (red and blue).  The minor variations in $q$ and $Q$ are the result of mild secular eccentric oscillations driven by the planets' mutual perturbations.}
% 	\label{fig:matt_fig}
% \end{figure}

\subsection{Modeling Transit Timing Variations}\label{sec:ttvs}
In addition to the transit modeling with the planets orbiting on linear ephemerides presented in Section~\ref{sec:transit}, we also generated a model that allowed the transit times to vary. This model enabled us to search for transit timing variations caused by either the two planets perturbing each other, or from a third body in the system. The model was very similar to the linear ephemeris model, except that (i) we used 20-s cadence data where available, (ii) we added a partial transit observed by the 1.6-m telescope at Observatoire du Mont-M\'egantic of TOI-2095 c (iii) we enabled the transit times to shift from the linear ephemeris, with a Gaussian prior with standard deviation of 0.03 days and mean of zero relative the a linear ephemeris, and (iv) we only allowed circular orbits. The reason for only allowing circular orbits here is that we are only interested in measuring the transit times, and neglecting non-circular orbits speeds up the calculation.

\begin{figure}
	\centering
	\includegraphics[width=0.6\textwidth]{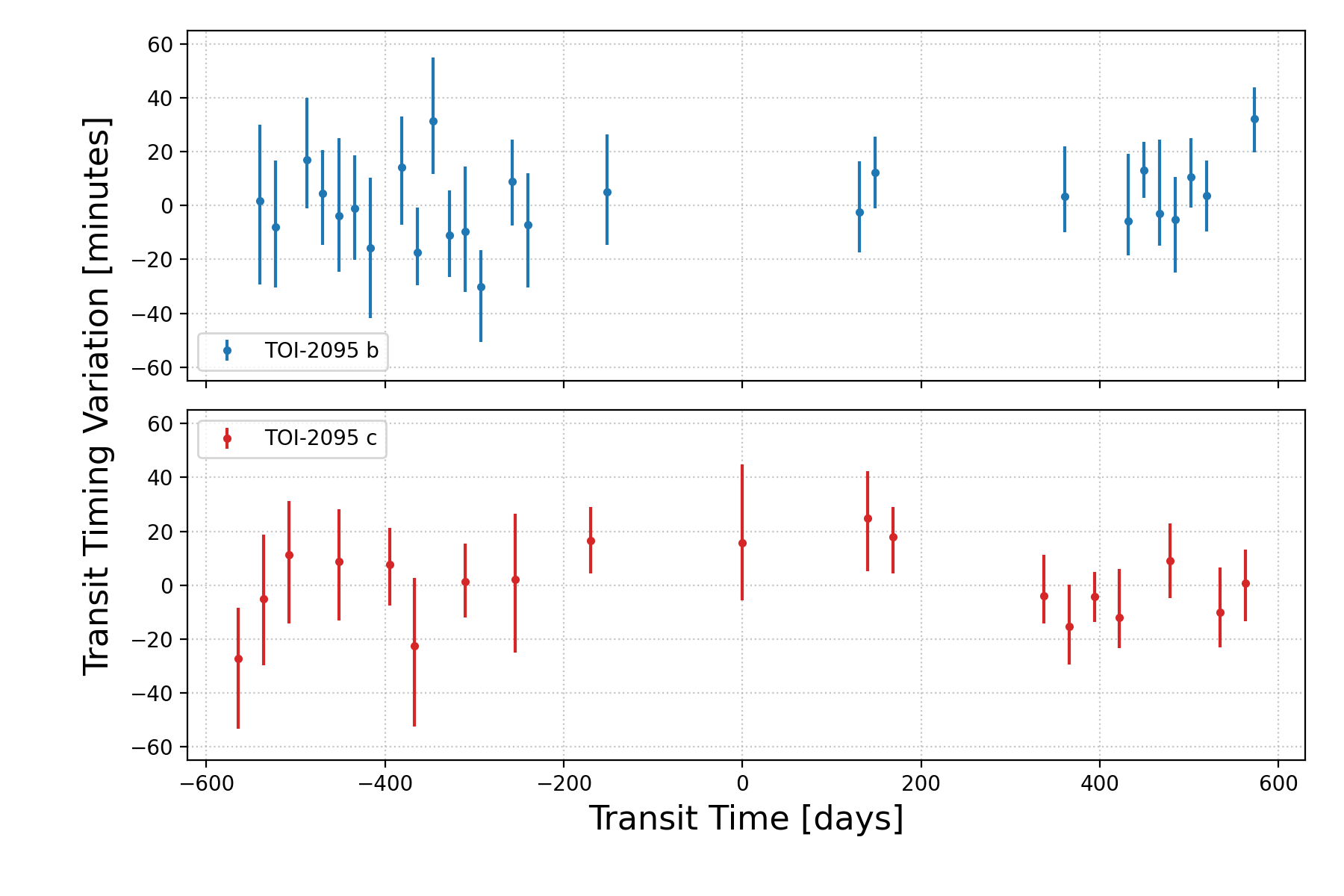}
	\caption{Observed minus calculated transit times for TOI-2095 b and c show no significant evidence for TTVs at this time. However, the number of transits is limited, and further observations may reveal a longer superperiod for transit times.}
	\label{fig:ttv}
\end{figure}

We did not detect significant transit timing variations for either planet over the 1170 day span that the observations cover. Figure~\ref{fig:ttv} shows the deviation from a linear ephemeris for the two planets. Planet b shows flat light TTV curves. However, the TTV model for TOI-2095 c shows hints of a turnover in the transit times (the first and last transits are earlier and the central transits occur later). This may be something to investigate further with additional TESS transit observations. 

\section{Discussion}

\subsection{Implications for Planet Formation}

A large fraction of the known super-Earth and sub-Neptune population reside in multiple planet systems that display a remarkable degree of intra-system uniformity \citep{Adams2019,Weiss2022}, in terms of planet masses, sizes, and circular/coplanar orbits. While there exists ongoing debate about the theories that support the so-called peas-in-a-pod phenomena for the population of compact multiplanet systems, the architecture of the TOI-2095 planetary system is consistent with these patterns. The TOI-2095 planet sizes are comparable, the period ratio of the TOI-2095 planets resides in the middle of the distribution of peas-in-a-pod systems, and the outer planet is slightly larger. Assuming this theory holds, we would expect the next outer planet (if one were to exist) to reside with a near 45 day period, which could motivate future transit searches if more data and a longer baseline are collected (which is feasible given TESS has been successful with extended missions).

The TOI-2095 system also provides an excellent laboratory to test formation mechanisms of the super-Earth and mini-Neptune populations. For close-in planets, photoevaporation and core-powered mass loss have been proposed as mechanisms that induce atmosphere loss and may explain this bimodality \citep[e.g.][]{Ginzburg2018,Gupta2019,Gupta2020,Owen2013,Owen2017,Cloutier2020}. However, small planets on wider orbits (greater than about 20 days around M dwarfs), such as TOI-2095 b (17.7 days) and TOI-2095 c (28.2 days), present an opportunity to distinguish between formation pathways \citep{Lee2022}. This is because at wide separations, the atmospheric mass loss timescales for small rocky planets often exceed the age of the planetary system. As a result, long-period rocky planets must have formed rocky and therefore are not being sculpted by atmospheric escape. This motivates follow-up mass measurements of the TOI-2095 planets to determine if they are rocky and to (potentially) rule out proposed formation mechanisms that rely on atmospheric escape.

\subsection{Prospects for Atmospheric Characterization}\label{sec:atmos}%\comment{Dana}

\begin{figure*}[ht!]
\plotone{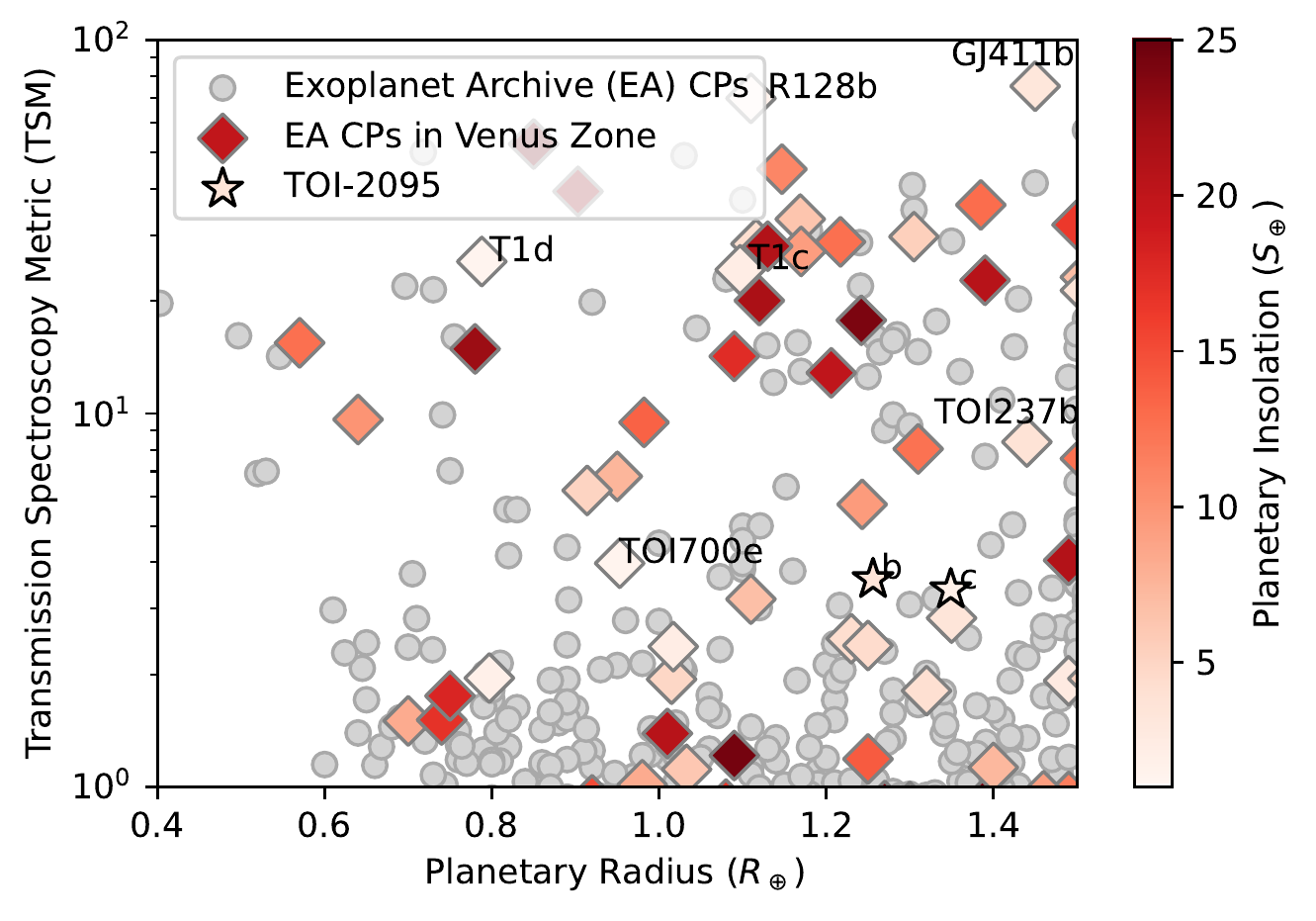}
\caption{Transmission Spectroscopy Metric (TSM) for the TOI-2095 planets, compared to the TSM for confirmed planets (CP) in the NASA Exoplanet Archive. The colorbar indicates other planets within the VZ. We label those planets near the outer edge of the VZ with equilibrium temperatures less than 400 K. Table \ref{tab:TSM_compare} lists the properties of the eight planets labeled in this figure. \label{fig:TSMcomparison}}
\end{figure*}

To determine the suitability of the TOI-2095 planets for atmospheric characterization, we computed the Transmission Spectroscopy Metric (TSM) and Emission Spectroscopy Metric \citep[ESM, ][]{kempton2018}, standardized metrics which are proportional to the expected transmission or emission S/N for a relevant \textit{JWST} observation, for the planets. We then compared those values to the TSM and ESM of all confirmed planets listed within the NASA Exoplanet Archive.\footnote{NASA Exoplanet Archive, \url{https://exoplanetarchive.ipac.caltech.edu/}, accessed on 1 November 2022.} Figure~\ref{fig:TSMcomparison} shows this comparison, with Venus Zone (VZ) planets depicted in color according to their insolation flux, and all other planets in gray. We followed the methodology of \citet{Kane2014} and \citet{ostberg2019} to define the boundaries of the VZ. Specifically, the inner edge of the VZ is set to 25$S_\oplus$, coinciding with the amount of flux that would place Venus on the Cosmic Shoreline, where the planet would start to experience severe atmospheric loss \citep{zahnle-catling2017}. We adopted the Runaway Greenhouse boundary as the outer edge of the VZ. The effective insolation flux at this boundary depends upon stellar type, and is defined by \cite{kopparapu2013, kopparapu2014} as

\begin{equation}\label{eq:VZouteredge}
    S_{\rm eff} = S_{\rm eff, \odot} + aT_{\star} + bT_{\star} ^2 + cT_{\star}^3 + dT_{\star}^4,
\end{equation}

\noindent where $T_{\star} = T_{\rm eff} - 5780$ K. \citet{kopparapu2013} provide the coefficients for stellar temperatures 2600 K $\le T_{\rm eff} \le$ 7200 K as $ S_{\rm eff, \odot} = 1.0512$, $a = 1.3242 \times 10^{-4}$, $b = 1.5418 \times 10^{-8}$, $c = -7.9895 \times 10^{-12}$, and $d = -1.8328 \times 10^{-15}$.

In order to estimate TSM values for TOI-2095 b and c, we estimated masses for the planets using the formula $1.436R_{p}^{1.7}$ \citep{louie2018,kempton2018}, which is based upon the \cite{Chen2017} mass-radius relationship. Following \cite{kempton2018}, we calculated planetary equilibrium temperature assuming zero albedo and uniform day-night heat redistribution. Using these values, we computed TSM to be 3.60 and 3.38 and ESM to be 0.29 and 0.15 for TOI-2095 b and TOI-2095 c, respectively. All other quantities required to calculate TSM and ESM for the TOI-2095 planets are taken from Tables \ref{table:hoststarinfo_derived} and \ref{tab:mcmc} of this work. In actuality, non-zero albedo and non-uniform heat redistribution can significantly impact the potential atmospheric composition and detectability and these assumptions should be carefully considered prior to any potential spectroscopic investigation. However, the TSM and ESM remain valuable tools to compare the potential observability of spectroscopic features between observation candidates in the absence of more in-depth knowledge.

In Table~\ref{tab:TSM_compare}, we compare the TSM values between the TOI-2095 planets and other VZ planets which have equilibrium temperatures less than 400 K. Although six confirmed VZ planets with $R_{\rm p} < 1.5R_\oplus$ have higher TSM values, the TOI-2095 planets offer observational advantages. % making them compelling targets to explore the atmospheres of transition radius planets near the outer edge of the VZ. 
Namely, the planets are within the TESS and JWST CVZs, making it easier to schedule multiple transit observations. The host star is relatively bright, yet not so bright that saturation may be an issue with most JWST near infrared instruments.\footnote{NIRSpec Prism mode generally saturates at $m_J \lesssim 10.5$, so only this instrument mode may present problems for the TOI-2095 system.} Finally, the lack of observed starspots and stellar flares indicates the star is relatively inactive. Stellar activity has been shown to complicate our interpretations of transmission spectroscopy observations \citep[e.g.,][]{rackham2018,Barclay2021,SAG212022}.

\begin{table}
\caption{Comparison of TOI-2095 to other VZ planets with $T_{\rm eq} < 400$ K}
\label{tab:TSM_compare}
\begin{tabular}{lccccccc}
Planet Name & TSM & R$_p$ ($R_\oplus$) & M$_p$ ($M_\oplus$) & R$_*$ ($R_\odot$) & T$_{\rm eq}$ (K) & m$_J$ (mag) & S$_p$ ($S_\oplus$) \\
 \hline
GJ 411 b & 75.259 & 1.450 & 2.690 & 0.370 & 350. & 4.320 & 3.13 \\
Ross 128 b & 69.839 & 1.110 & 1.400 & 0.200 & 301. & 6.505 & 1.38 \\
TRAPPIST-1 d & 25.688 & 0.788 & 0.388 & 0.120 & 288. & 11.354 & 1.11 \\
TRAPPIST-1 c & 24.406 & 1.097 & 1.308 & 0.120 & 342. & 11.354 & 2.21 \\
TOI-237 b & 8.389 & 1.440 & 2.670 & 0.210 & 388. & 11.740 & 3.70 \\
TOI-700 e & 3.972 & 0.953 & 0.818 & 0.420 & 273. & 9.469 & 1.27 \\
TOI-2095 b & 3.598 & 1.256 & 2.116 & 0.451 & 375. & 9.797 & 3.26 \\
TOI-2095 c & 3.377 & 1.349 & 2.389 & 0.451 & 320. & 9.797 & 1.75 \\
\end{tabular}
\end{table}

\subsection{Venus Analogs}\label{sec:venus}

Recent studies of planetary habitability have emphasized the need to leverage the limited data inventory of terrestrial atmospheres from within the solar system (i.e. of Earth, Venus, and Mars) \citep{Kane2021b}. In particular, understanding the atmospheric and interior evolution of Venus is considered critical within the context of planetary habitability and as a parallel to an Earth-based climate model \citep{Popp2016,Kane2019,Margot2021,Kane2022}. Models of early Venus suggest that water may never have condensed on the surface due to an extended magma phase \citep{Hamano2013} and/or cloud formation on the night-side of the planet \citep{Turbet2021}. Alternatively, for scenarios in which surface water condensation occurred, Venus may have maintained temperate surface conditions for several billion years, enabled by cloud formation at the sub-stellar point \citep{Way2016,Way2020}. In fact, it has been shown that both a habitable and waterless past for Venus self-consistently reproduce modern bulk atmospheric composition, inferred surface heat flow, and observed $^{40}$Ar and $^4$He \citep{Krissansen-Totton2021}, further underscoring the need for additional investigations. In addition, early orbital dynamic effects may have enhanced water loss from the young Venus \citep{Kane2020}, an effect that may have a more pronounced influence for eccentric exoplanets \citep{Barnes2013,Palubski2020}. Venus also serves as a local laboratory for atmospheric loss effects, with application to exoplanets \citep{Dong2020}. A detailed investigation of these various facets of our sister planet required significantly more planetary data, which has motivated further Venus missions over the coming decade \citep{Garvin2022,Smrekar2022,Ghail2020}.

Numerous discovered exoplanets have been proposed as potential exoVenus candidates, including Kepler-69 c \citep{Kane2013}, Kepler-1649 b \citep{Angelo2017,Kane2018,Kane2021a}, TRAPPIST-1 c \citep{Lincowski2018}, and GJ~3929 b \citep{Beard2022}. Indeed, the large number of close-in exoplanets has enabled a statistical consideration of Venus Zone planet occurrence rates \citep{Kane2014}, along with suitable targets for atmospheric follow-up observations \citep{ostberg2019,Lincowski2019,Yaeger2019c,Colby2023}. Such follow-up work requires a detailed knowledge of the Venusian atmospheric chemistry and structure, and how these details manifest in the expected spectral signatures that are acquired \citep{Schaefer2011,Ehrenreich2012,Barstow2016,Jordan2021}. As described in Section~\ref{sec:atmos} and shown in Figure~\ref{fig:TSMcomparison}, the TOI-2095 planets fall alongside numerous other interesting exoVenus candidates and present additional prospects for studying terrestrial atmospheric evolution as a function of such aspects as planetary radius and insolation flux.

Currently, the primary method for studying the atmospheric composition of Venus-like worlds will be through transmission spectroscopy, which is used to determine the wavelengths at which light is absorbed when passing through a planet’s atmosphere. Venus’ transmission spectrum was modeled in preparation for the transit of Venus in 2012, which demonstrated that the Venusian cloud and haze layers prevent transmission spectroscopy from probing the atmosphere below an altitude of 80~km \citep{Ehrenreich2012}. \citet{Lincowski2018} modeled the transmission spectra of the TRAPPIST-1 planets assuming they had 10-bar as well as 92.1 bar (the surface pressure of Venus) Venus-like atmospheres. In both cases, their work illustrated that the weaker CO$_2$ absorption bands at 1.05 and 1.3 $\mu$m and absorption caused by sulfuric acid clouds are likely to be the best avenues for determining if a planet has a Venus-like atmosphere and may help constrain a high CO$_2$ abundance. Simulated JWST observations of the TRAPPIST-1 planets with Venus-like atmospheres showed that their atmospheres could be detected in less than 20 transit observations, but discerning their compositions would take more than 60 transit observations \citep{Yaeger2019}. We adopted a similar approach and estimated the transmission spectrum of TOI-2095 b with a cloudless, 10 bar Venus-like atmosphere using the Planetary Spectrum Generator (PSG) \citep{villanueva2018}. %(Figure \ref{fig:Spectra}; upper plot). 
We simulated 100 JWST transit observations of this hypothetical TOI-2095 b Venus, and while CO$_2$ absorption features could be seen at 2.0, 2.7, and 4.3~$\mu$m, these features were only a few ppm in depth, which is far too small to be detected by JWST. Better constraints on the viability of atmospheric studies with the TOI-2095 system can be provided if mass measurements are obtained. 

\begin{figure*}
    \centering
    \includegraphics[width=0.85\textwidth]{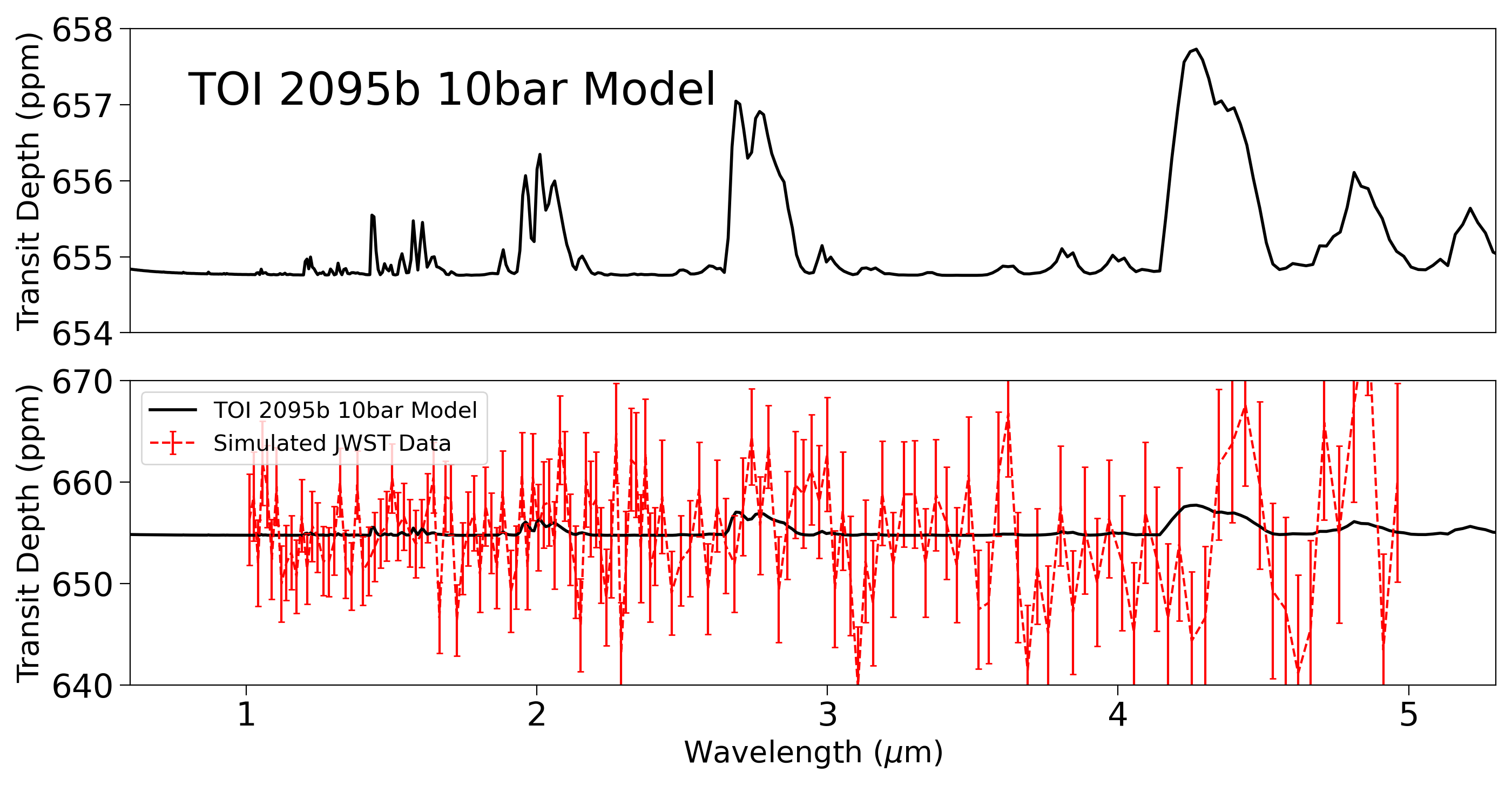}
    \caption{An estimated transmission spectrum of TOI-2095 b with a 10 bar, cloudless Venus-like atmosphere (upper plot), and simulated JWST data of the same planet assuming 100 transit observations (lower plot). Note that the black line is the same spectrum in both the upper and lower plots, but the scale of the y-axis is larger in the lower plot to capture the uncertainty in the simulated JWST data.
    \label{fig:Spectra}}
\end{figure*}

\section{Summary and Conclusions} %\comment{Elisa}
The TOI-2095 two-planet system provides another valuable system discovered by TESS that is amenable to follow-up observations that can place constraints on the system's bulk composition and formation history. The star, a 0.47 solar mass M1V dwarf, lies in the TESS continuous viewing zone, and we present results based on 24 TESS Sectors. 

This multiplanet system is dynamically stable, and no signs of transit timing variations have been observed thus far (although a longer baseline of additional TESS observations could reveal TTVs). Amongst the most exciting prospects for characterizing the TOI-2095 system is the potential for obtaining precise radial velocity mass measurements. With relatively wide orbits, establishing the composition and determining whether either planet is indeed rocky will provide valuable clues into the formation mechanisms that sculpt the widely-studied, but still widely-debated, problem of why a radius valley exists between rocky planets and those with H/He envelopes.   

The TOI-2095 planets are only about 30\% larger than Earth, and have insolation values between 1.7 and 3.2 times that which Earth receives from the Sun, placing them in the Venus-class regime. We explored the feasibility for transmission and emission spectroscopic measurements to probe their atmospheres via missions like JWST. While the calculated metrics (TSM and ESM) that indicate their potential for such measurements rank lower than six other small exoplanets in the Venus-class regime, TOI-2095 is a relatively quiet star which is beneficial for interpreting atmospheric spectra. We simulated the transmission spectrum of TOI-2095 b assuming a cloudless, 10 bar Venus-like atmosphere, and found that CO$_2$ absorption features would be far too small to be detected by JWST. As we obtain more data on the atmospheres and compositions of Venus zone planets (such as the JWST observations of the TRAPPIST1 planets), we will gain a better understanding on the assumptions and range of inputs that can better simulate and model this class of planets, as well as the key observables we should look for in high-precision JWST data that can place these small planets into context.

The field of transiting exoplanet science is bright, and missions like TESS, PLATO, and the upcoming Nancy Grace Roman Space Telescope have potential to deliver many thousands of additional exoplanets to study \citep{montet2017,Barclay2018}. Systems that orbit bright stars and are amenable to follow-up studies, like TOI-2095, offer an excellent opportunity to shed light on composition and formation theories, and ultimately identify trends that will set the stage for the interpretation of exoplanet populations revealed by future missions.

\begin{acknowledgments}

We are thankful for support from GSFC Sellers Exoplanet Environments Collaboration (SEEC), which is funded by the NASA Planetary Science Division’s Internal Scientist Funding Model. B.J.H. acknowledges support from the Future Investigators in NASA Earth and Space Science and Technology (FINESST) program grant 80NSSC20K1551. T.A.B.'s and D.R.L.'s research activities were supported by an appointment to the NASA Postdoctoral Program at the NASA Goddard Space Flight Center, administered by Oak Ridge Associated Universities under contract with NASA. KAC acknowledges support from the TESS mission via subaward s3449 from MIT.

This paper includes data collected by the TESS mission, which are publicly available from the Mikulski Archive for Space Telescopes (MAST). Funding for the TESS mission is provided by NASA's Science Mission Directorate. We acknowledge the use of public TESS Alert data from pipelines at the TESS Science Office and at the TESS Science Processing Operations Center. Resources supporting this work were provided by the NASA High-End Computing (HEC) Program through the NASA Advanced Supercomputing (NAS) Division at Ames Research Center for the production of the SPOC data products. This research has made use of the Exoplanet Follow-up Observation Program website, which is operated by the California Institute of Technology, under contract with the National Aeronautics and Space Administration under the Exoplanet Exploration Program. This work has made use of data from the European Space Agency (ESA) mission {\it Gaia} (\url{https://www.cosmos.esa.int/gaia}), processed by the {\it Gaia} Data Processing and Analysis Consortium (DPAC, \url{https://www.cosmos.esa.int/web/gaia/dpac/consortium}). Data presented herein were obtained at the W. M. Keck Observatory, which is operated as a scientific partnership among the California Institute of Technology, the University of California and the National Aeronautics and Space Administration. The Observatory was made possible by the generous financial support of the W. M. Keck Foundation. The authors acknowledge the USNA Advanced Research Computing Support (ARCS) office (\url{https://www.usna.edu/ARCS/}) which was made available for conducting the research reported in this paper. The material is based upon work supported by NASA under award number 80GSFC21M0002. This work makes use of observations from the \lco\ network. Part of the \lco\ telescope time was granted by NOIRLab through the Mid-Scale Innovations Program (MSIP). MSIP is funded by NSF.

This research has made use of the NASA Exoplanet Archive, which is operated by the California Institute of Technology, under contract with the National Aeronautics and Space Administration under the Exoplanet Exploration Program. This research has made use of the Exoplanet Follow-up Observation Program (ExoFOP; DOI: 10.26134/ExoFOP5) website, which is operated by the California Institute of Technology, under contract with the National Aeronautics and Space Administration under the Exoplanet Exploration Program.

\end{acknowledgments}

\vspace{5mm}
\facilities{TESS, \lco\ (Sinistro), OMM:1.6, Keck:II (NIRC2), FLWO:1.5m (TRES), Exoplanet Archive, ExoFOP, Gaia}

\software{
Arviz \citep{exoplanet:arviz},
AstroImageJ \citep{Collins:2017},
astropy \citep{exoplanet:astropy13,exoplanet:astropy18}, 
celerite2 \citep{exoplanet:foremanmackey17, exoplanet:foremanmackey18}, 
exoplanet \citep{exoplanet:exoplanet}, 
DAVE \citep{Kostov2019}, 
% Forecaster \citep{forecaster}, 
IPython \citep{ipython}, 
Jupyter \citep{jupyter}, 
Lightkurve \citep{lightkurve}, 
% M\_-M\_K- \citep{mann19}, 
Matplotlib \citep{matplotlib},
Mercury6 \citep{chambers99},
NumPy \citep{numpy}, 
Pandas \citep{pandas}, 
PyMC3 \citep{exoplanet:pymc3}, 
% SciPy \citep{scipy}, 
% stardate \citep{stardate,Angus2019}, 
STARRY \citep{exoplanet:luger18, exoplanet:agol20}, 
Tapir\citep{Jensen:2013}, 
% TRANSITFIT5 \citep{Rowe2015,Rowe2016c},
Theano \citep{exoplanet:theano}, 
TRICERATOPS \citep{giacalone2020vetting},
% xoflares \citep{xoflares}
% TTVFast \citep{Deck2014}, 
% TTV2Fast2Furious \citep{Hadden2018}, 
vespa \citep{Morton2012,vespa}
}

\bibliography{references}{}
\bibliographystyle{aasjournal}

\end{document}